\pdfoutput=1 

\RequirePackage{fix-cm}
\documentclass{svjour3}                     
\smartqed  
\usepackage{graphicx}
\journalname{The Astronomy and Astrophysics Review}
\usepackage{natbib}
\usepackage[usenames,dvipsnames]{color}
\bibpunct{(}{)}{;}{a}{}{,}

\newcommand\kms{km\,s$^{-1}$}
\newcommand\smyr{M$_\odot$\,yr$^{-1}$}
\newcommand\taug{{\rm \tau_g}}

\begin{document}

\title{Star formation sustained by gas accretion
}


\author{Jorge~S\'anchez~Almeida \and Bruce~G.~Elmegreen \and \\Casiana~Mu\~noz-Tu\~n\'on \and Debra~Meloy~Elmegreen
}

\authorrunning{J. S\'anchez~Almeida} 

\institute{
J.~S\'anchez~Almeida \at
Instituto de Astrof\'\i sica de Canarias, E-38205 La Laguna, Tenerife, Spain\\
Departamento de Astrof\'\i sica, Universidad de La Laguna, Tenerife, Spain\\
\email{jos@iac.es}             
\and
B.~G.~Elmegreen \at 
IBM Research Division, T.J. Watson Research Center, Yorktown Heights, NY 10598, USA\\
\email{bge@us.ibm.com}
\and
C.~Mu\~noz-Tu\~n\'on \at
Instituto de Astrof\'\i sica de Canarias, E-38205 La Laguna, Tenerife, Spain\\
Departamento de Astrof\'\i sica, Universidad de La Laguna, Tenerife, Spain\\
\email{cmt@iac.es}
\and 
D.~M.~Elmegreen \at
Department of Physics and Astronomy, Vassar College,  Poughkeepsie, NY 12604, USA\\
\email{elmegreen@vassar.edu}
}

\date{\today}

\maketitle

\begin{abstract}
Numerical simulations predict that  metal-poor gas accretion 
from  the cosmic web fuels the formation of disk galaxies.
This paper discusses how cosmic gas accretion controls 
star formation, and summarizes the physical properties 
expected for the cosmic gas accreted by galaxies. The paper
also collects observational evidence for gas accretion 
sustaining star formation. It reviews evidence inferred from 
neutral and ionized hydrogen, as well as from stars. A number of properties 
characterizing large samples of  star-forming galaxies can be explained 
by metal-poor gas accretion, in particular, the relationship between 
stellar mass, metallicity, and star formation rate (the so-called fundamental 
metallicity relationship). They are put forward and analyzed. 
Theory predicts gas accretion to be particularly important at high 
redshift, so indications based on distant objects  are reviewed, 
including the global star formation history of the universe, and the 
gas  around galaxies as inferred from absorption features 
in the spectra of background sources. 
%
\end{abstract}

%

\section{Introduction}\label{introduction}

\begin{figure}
\includegraphics[width=1.0\linewidth]{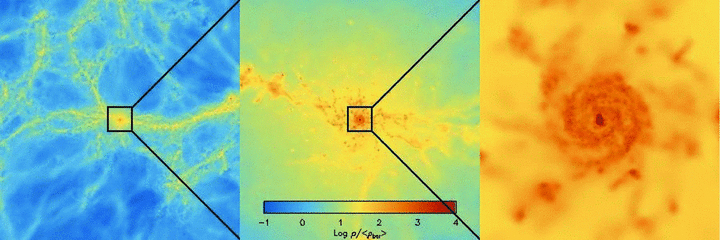}
\caption{
Zoom into a simulated halo  at $z=2$ with halo mass
${\rm M_{halo}\simeq 10^{12}\,M_\odot}$  \citep[from][]{2010MNRAS.402.1536S}.
From left to right, the images are 10, 1 and 0.1~Mpc on a side. All slices are 1~Mpc thick. 
The color coding shows the projected gas density in  logarithmic scale, 
with the equivalence given in the inset.  Densities are referred to the mean density of
 the universe at the time.  
}
\label{just_nice}
\end{figure}

Numerical simulations predict that accretion of metal-poor gas from 
the cosmic web fuels  the formation of disk galaxies 
(e.g.,
\citeauthor{2006MNRAS.368....2D}~\citeyear{2006MNRAS.368....2D},
\citeauthor{2009Natur.457..451D}~\citeyear{2009Natur.457..451D},
\citeauthor{2012RAA....12..917S}~\citeyear{2012RAA....12..917S},
\citeauthor{2012ApJ...745...11G}~\citeyear{2012ApJ...745...11G};
see also Fig.~\ref{just_nice}).
The gas originally resides outside the virial radius of the dark
matter (DM) halo that hosts the galaxy, and by accretion over cosmic 
time it  becomes part of the pool of baryons that forms 
new stars. This cosmological gas supply has a strong dependence on 
redshift and halo mass. 
When the gas  encounters a massive halo  ($> 10^{12} {\rm M}_\odot$), 
it becomes shock heated and requires a long time to cool and settle into 
the galaxy disk. For less massive haloes,  cool gas streams can 
reach the inner halo or disk directly. 
This so-called  cold-flow accretion may have an extreme impact on the disk, 
or it may  provide a gentle gas supply that is transported  radially in the disk.
Since high redshift haloes tend to be low in mass, cold-flow accretion 
is predicted to be the main mode of galaxy growth in early times. Ultimately, the galaxies 
evolve into  a quasi-stationary  state (Sect.~\ref{stationary_state}), 
where inflows and outflows balance the star formation rate 
(SFR), a phase that still goes on for most of them.

The importance of  gas infall is as clear from numerical simulations 
as it is obscure to  observations. 
This paper gives an overview of this rapidly evolving field, emphasizing the role 
of metal-poor gas accretion  to sustain star formation in the local universe. 
We limit ourselves to the global picture, leaving aside details about 
star formation processes 
\citep{2012ApJ...745...69K,2014ApJ...787L...7G}, 
stellar and active galactic nucleus (AGN) feedback
\citep{2010ApJ...711...25S,2013arXiv1311.2073H,2013arXiv1311.2910T},
secular evolution  
\citep{2013NewAR..57...29B,2013seg..book....1K},
dense cluster environments 
\citep{2011JPhCS.280a2007S,2012ARA&A..50..353K},
and the growth of black holes (BH) through cosmic gas accretion 
\citep{2011A&A...535A..72H,2013ApJ...773....3C}. 
Other recent reviews covering cosmic gas accretion
from different perspectives are in
\citet{2008A&ARv..15..189S}, 
\citet{2012RAA....12..917S},
\citet{2014ASPC..480..211C},
\citet{2014IAUS..298..228F},
\citet{2013MNRAS.430.1051C} (observationally oriented),
\citet{2013seg..book..555S},
\citet{2010PhR...495...33B} (theoretically oriented),
and 
\citet{2014arXiv1403.0007M} (emphasizing the high redshift aspects).


In addition to gas accretion, galaxies also grow through  mergers 
\citep[e.g,][]{2011MNRAS.413..101G,2013seg..book....1K}. 
Simulations suggest that direct accretion  from the cosmic web dominates mergers  
by about an order of  magnitude 
\citep[e.g.,][]{2011MNRAS.413.1373W,2012A&A...544A..68L,2013MSAIS..25...45C,2011MNRAS.414.2458V}.
For example, 
the Aquarius project to study galaxy formation in ${\rm \Lambda}$ cold dark matter
(${\rm \Lambda}$CDM) haloes shows inside-out growth, 
with major mergers contributing  less than 20\,\%\ to the total mass growth 
\citep{2011MNRAS.413.1373W}. These simulations indicate that most of the baryons from which visible
 galaxies form are accreted diffusely, rather than through mergers, and only relatively rare major mergers
affect galaxy structure at later times. 
\citet{2012A&A...544A..68L} use a multi-zoom simulation to 
quantify mass assembly and find that, on average,  77\,\% of growth is from smooth accretion, with 
23\,\% from mergers. In dense regions, mergers play a more  prominent role,  although even there 
gas accretion still dominates the growth in the \citeauthor{2012A&A...544A..68L} simulations. 
Groups and clusters are the primary environments in which mergers are 
important. 
Since our review paper is focused on gas accretion driving star formation (SF),
we do not consider dense environments 
but concentrate on the evolution of galaxies in relative isolation. 
For the same reason, we ignore satellite galaxies where ram pressure stripping 
and  starvation are important.

%
%


This paper is organized as follows: Sect.~\ref{theory} summarizes the 
expected properties of the cosmic gas that fuels SF.
Section~\ref{stationary_state} 
puts  forward a simple analytical model that intuitively explains how  
gas accretion controls SF.  Section~\ref{physical_picture} describes the physical properties 
of accreting gas  in numerical simulations, with outflows 
discussed in Sec.~\ref{winds} and  disk growth  in  Sect.~\ref{disks}.
Section~\ref{gas_accretion} reviews the evidence for gas accretion inferred from 
neutral (Sect.~\ref{neutral_gas}) and ionized hydrogen (Sect.~\ref{hiiregions}).
Evidence from stellar properties is included in Sect.~\ref{star_accretion}. 
A number of observational properties characterizing large samples of 
star-forming galaxies can be explained if the SF is driven by metal-poor gas accretion.
These properties are put forward and discussed in Sect.~\ref{scaling_law}, specifically, 
the stellar mass-metallicity-SFR relationship (Sect.~\ref{fmmr}),
the stellar mass-metallicity-gas mass relationship (Sect.~\ref{fmr2}),
and the stellar mass-metallicity-size relationship (Sect.~\ref{fmr3}). 
Theory predicts  gas accretion to be particularly important at high redshift.
Section ~\ref{redshift} reviews observational evidence for accretion at 
high redshift, starting with the SF history of the universe (Sect.~\ref{sfhu}). 
In addition, it treats the (secondary) role of mergers 
(Sect.~\ref{role_merger}), the evidence for gas around galaxies as 
inferred from absorption  features in the spectra of background sources 
(Sect.~\ref{cwabs}), and the imaging of web gas through its emission
(Sect.~\ref{cosmic_emission}). 
Theory predicts that most of the gas going into stars
in a typical SF episode is not recycled gas from previous star formation, 
but it comes from accretion.
Section~\ref{myfraction} collects the few measurements of the 
fraction of SF produced by gas accretion available so far. 
The review concludes by summarizing the role of gas accretion 
in star formation, and indicating several open issues to be explored in the
future (Sect.~\ref{conclusions}).
The acronyms and main symbols used along the text are listed in Table~\ref{acronyms}.


\section{Cosmological accretion: basics}\label{theory}

During the expansion of the Universe, cosmological gas that falls into the
potential well of a DM concentration eventually shocks when it meets
other gas and stops or deflects in its path \citep{1977MNRAS.179..541R,
1978MNRAS.183..341W, 1977ApJ...215..483B, 1977ApJ...211..638S}.  
The post-shock temperature increases with the speed of the fall and 
therefore with the depth of the potential well, and this depth scales 
with the halo mass. 
At small enough halo mass, around  $10^{12}\;{\rm M_\odot}$ for a modest metal 
abundance \citep{2003MNRAS.345..349B}, the temperature is sufficiently low that the 
post-shock cooling time becomes short compared to the dynamical time. Then  
cooling behind the shock efficiently removes its pressure support and the shock 
collapses. This collapse is the result of an instability that occurs at a 
low effective adiabatic index, when the pressure increase following compression 
cannot offset  the increase in self-gravity. If the shock collapses completely, 
then the gas may not be heated to the halo virial temperature until it mixes with the 
inner regions or hits the disk. 
The structures falling in are expected to take the form of filamentary streams
or clouds \cite[e.g.,][]{2005MNRAS.363....2K,2009Natur.457..451D}.
The importance of such cold streams  is that the gas can reach 
the disk faster, 
so SF can begin sooner and achieve a rate that is within a factor of a
few of the baryonic accretion rate in the halo  \citep[e.g.,][]{2009ApJ...694..396B}.

\begin{figure}
\includegraphics[width=0.6\linewidth]{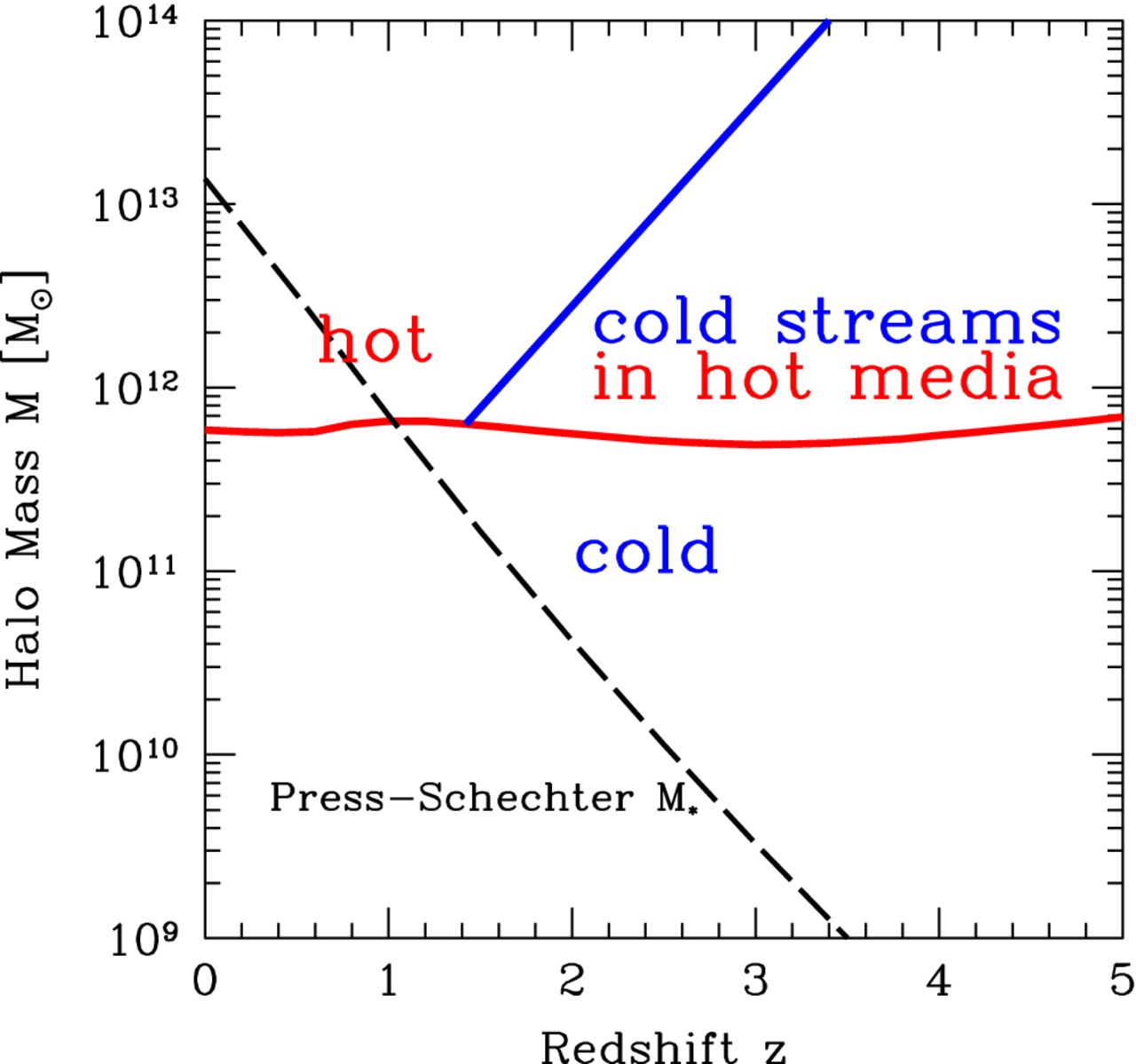}
\caption{
Predicted penetration of cold gas streams into the halo centre as a 
function of halo mass and redshift. 
The nearly horizontal curve is the  threshold mass below which the
flows  are predominantly cold and above which a shock-heated medium is
present. The inclined solid curve is the upper limit for cold 
streams in a hot medium to be present. Above this line the gas is 
all shock heated shutting off the gas supply for star formation.
This schematic  is based on analytic  spherically-symmetric 
calculations by  \citet{2006MNRAS.368....2D} and has been taken
from   \citet{2013MNRAS.435..999D}.
The slanted dashed line is the critical mass in the Press-Schechter
formalism, above which structures form.
}
\label{dekelf10}
\end{figure}

At early times, when galaxies were generally low-mass, and at recent
times in the case of low-mass galaxies, 
a significant fraction of the  gas from cosmological
accretion remains cold and falls directly to the center  
\citep{2003MNRAS.345..349B,
2005A&A...441...55S, 2011ApJ...735L...1S, 2011MNRAS.414.2458V}. 
Figure~\ref{dekelf10} shows  when the cold and  hot accretion regimes operate 
as a function of halo mass and redshift \citep[from][]{2013MNRAS.435..999D}. 
Haloes below  $\sim 10^{12}\;{\rm M_\odot}$ are dominated
by cold accretion at all redshifts, whereas even high mass galaxies at $z>1$ could have 
some cold accretion in streams that  penetrate the surrounding hot haloes 
\citep[see also ][]{2008MNRAS.390.1326O, 2009Natur.457..451D, 2009MNRAS.395..160K,2011ApJ...735L...1S}. 

Much progress has been made on the details of galaxy accretion since the
early studies, with few changes to the basic conclusion that the cold fraction of
accretion is determined by the dark halo mass. This section summarizes the physical 
properties of the gas that  falls in and forms stars as predicted by recent numerical 
simulations.  We outline expected observational properties, keeping in mind that 
the numerical methods used to carry out the simulations may matter for some 
important details since the resolution is not good  enough to capture many 
of the  physical processes controlling the  gas infall
\citep[the typical mass resolution is about
$10^6$-$10^7{\rm \,M_\odot}$;][] {2009ApJ...694..396B,2012MNRAS.423.2991V}.
For example, the moving mesh code AREPO gets less
purely-cold gas at high halo mass than do
smoothed particle hydrodynamics (SPH) codes because cold 
filaments break apart and mix with hot halo gas more readily in AREPO
\citep{2013MNRAS.429.3353N}.
Similarly, \cite{2012MNRAS.425.2027K}, \cite{2012MNRAS.425.3024V} and
\cite{2012MNRAS.424.2999S} found higher star formation rates 
with AREPO than SPH because realistic turbulent cascades in AREPO 
channel the energy to denser scales where it dissipates at a greater rate.
Often simulations do not reach readshift $z=0$ and the predictions for the 
local universe come from higher $z$, considering that the accretion 
decreases with time.

%
\subsection{Simple analytical description}\label{stationary_state}

\paragraph{Quasi-stationary state.}
Numerical simulations produce galaxies that are originally very dynamic,
but as time goes on they enter a quasi-stationary phase were
inflows and outflows  balance  the SFR  
\citep[][]{
2008MNRAS.385.2181F,
2010MNRAS.402.1536S,
2012MNRAS.426.2166F,
2012MNRAS.421...98D,
2013MNRAS.435..999D,
2013MNRAS.433.1425B,
2013MNRAS.433.1910F,
2013MNRAS.436.2689A,
2014MNRAS.438.1552F}.
Such equilibrium follows naturally from the gas consumption timescale 
$\taug$ being shorter than the other relevant timescales in the problem.
For example, in the local universe $\taug$ is on the order of 1-2\,Gyr 
whereas the timescale for gas inflow from the web is some 7-8\,Gyr  
\citep[e.g.,][]{2013ApJ...772..119L}. Under this condition the gas inflow rate  determines 
both the SFR and the mass of gas in the galaxy \citep[][]{2013ApJ...772..119L}. 
The latter is just the gas mass needed to maintain the SFR given
the gas consumption time-scale. These conclusions
follow from mass conservation,
and can be summarized in a few analytical expressions which 
provide insight to understand the underlying physics.

Using equations from  chemical evolution models \citep[e.g.,][]{1990MNRAS.246..678E},  
the variation with time $t$ of the gas mass available to form stars ${\rm \dot{M}_g}$ 
is given by,
\begin{equation}
{{d{\rm M_g}}\over{dt}}\equiv {\rm \dot{M}_g}
=-(1-R)\,{\rm SFR}+{\rm \dot{M}_{in}}-{\rm \dot{M}_{out}},
\label{1steq}
\end{equation}
 which considers the formation of stars 
(1st term on the right-hand side of the equation), the
gas inflow rate ${\rm \dot{M}_{in}}(t)$, and the gas outflow rate
${\rm \dot{M}_{out}}(t)$. The symbol
$R$ in Eq.~(\ref{1steq}) stands for the fraction of the stellar mass
that returns to the interstellar medium (ISM) rather than being locked 
into stars and stellar remnants.
For simplicity, the SFR produced by a given mass 
of gas is parameterized in terms of the
star formation efficiency $\epsilon$, or its inverse
the gas consumption  time-scale $\taug$,
\begin{equation}
{\rm SFR}=\epsilon\, {\rm M_g}={{\rm M_g}\over{\taug}}.
\label{kslaw}
\end{equation}
The outflow rate is reasonably assumed to scale with the SFR,
\begin{equation}
{\rm \dot{M}_{out}}(t)=w\,{\rm SFR}(t),
\label{mass_load}
\end{equation}
with $w$ the so-called mass-loading factor.
Provided that all scaling factors $R, w$ and $\taug$ are constant,
the general solution of Eq.~(\ref{1steq}) is 
\begin{equation}
{\rm SFR}(t)={\rm SFR}(0)\,{\rm e}^{-t/\tau_{\rm in}}+
\int_0^t\,{\rm \dot{M}_{in}}(t^\prime)\,{\rm e}^{-(t-t^\prime)/\tau_{\rm in}}\,dt^\prime/\taug,
\label{2ndeq}
\end{equation}
\begin{displaymath}
\tau_{\rm in}=\taug\,/(1-R+w).
\end{displaymath}
Under the current assumption that $\taug$ is much
shorter than the other relevant timescales, 
$t \gg \tau_{\rm in}$, and the source term 
${\rm \dot{M}_{in}}(t^\prime)$ only contributes to the integral in Eq.~(\ref{2ndeq})
when $t^\prime\sim t$.
When this stationary-state is reached,
Eq.~(\ref{2ndeq}) becomes
\begin{equation}
{\rm SFR}(t)\simeq (1-R+w)^{-1}\,{\rm \dot{M}_{in}}(t),
\label{eqsfr}
\end{equation}
showing that the SFR is set by the gas infall rate (corrected for
returned fraction $R$ and outflows $w$). 
In addition, the mass of gas is also set by the infall rate 
to be the amount needed to maintain the SFR forced by the infall rate, 
\begin{equation}
{\rm M_g}(t)\simeq \taug\, {\rm SFR}(t)\simeq 
{{\taug}\over{1-R+w}}\,{\rm \dot{M}_{in}}(t).
\label{mykslaw}
\end{equation}
Equations~(\ref{eqsfr}) and (\ref{mykslaw}) tell us that 
galaxies self-regulate to maintain the gas needed to 
produce a SFR set by the balance between inflows and outflows.
We note that Eq.~(\ref{kslaw}), and so Eq.~(\ref{mykslaw}), is a version of the 
Kennicutt-Schmidt (KS) relation \citep[][]{1959ApJ...129..243S,1998ApJ...498..541K}
stating that the SFR scales as a power of the gas mass, with 
power index close to one. The details of this relation are not as important on the 
scale of a whole galaxy as they are for kpc-scales, where observations still have 
some uncertainties \citep[see the review in][]{2012ARA&A..50..531K}.
Observations suggest that $\taug$ goes from 0.5 to 2\, Gyr for galaxies 
in the redshift range  between 2 and 0  
\citep[e.g.,][]{2010MNRAS.407.2091G,2014ApJ...787L...7G}. 
Theory 
suggests that $\taug$ scales with the instantaneous Hubble time 
$t_H$ as $\taug\simeq 0.17\,t_H$ \citep{2013MNRAS.435..999D}.

In this stationary-state, a significant part of the SF relies on accreted 
gas because the gas in the disk that has been returned from previous stars (the 
returned fraction $R$) is not massive enough to maintain the level of SF very long.
The fraction of SF that comes from gas that has never been 
processed by stars in the galaxy 
can be determined by dividing the gas into one component of this type, 
${\rm M_{gf}}$, and another component that has been 
inside a star, ${\rm M_{gp}}$, 
\begin{equation} 
{\rm M_g = M_{gf}+M_{gp}}, 
\end{equation} 
where the subscripts f and p refer to fresh and processed gas, respectively.
The fresh gas follows an equation identical to Eq.~(\ref{1steq}) but 
without the returned fraction $R$, since the gas returned by stars is not fresh gas, 
\begin{equation} 
{\rm \dot{M}_{gf}} 
={\rm \dot{M}_{in}}-{\rm M_{gf}}/\taug\,-w\,{\rm {M}_{gf}}/\taug\,. 
\label{eqpristine} 
\end{equation} 
This equation states that all of the cosmic accretion adds to the fresh gas 
reservoir, which is depleted by star formation (both the locked-up mass and the 
returned mass) and by winds that are proportional to the SFR. 
We have assumed that winds carry away fresh gas in proportion to its mass, 
because the fresh and processed components should mix before significant star 
formation begins in them, and then both get dragged out by mass loading in the 
wind. 
These are reasonable assumptions since the 
time during which the gas is able to mix before star formation occurs in it, on 
average, is the consumption time, $\taug$, and this is generally much longer 
than the turbulent crossing time in the ISM.  There could be some situations, 
however, where fresh gas is not well mixed with processed gas before it gets 
into a star, and then the young star metallicity in those regions could be less 
than the average in the ISM. 
The evolution of ${\rm M_{gp}}$ is given by, 
\begin{equation} 
{\rm \dot{M}_{gp}} 
={\it R}\,{\rm M_{g}}/\taug-{\rm M_{gp}}/\taug\,-w\,{\rm {M}_{gp}}/\taug , 
\label{nonpristine} 
\end{equation} 
showing that the only source of processed gas is the mass that has been returned 
from star formation, while the loss of ${\rm M_{gp}}$ is from star 
formation and winds, as for the fresh gas loss. Since Eq.~(\ref{eqpristine}) is 
formally identical to Eq.~(\ref{1steq}), it admits a solution like 
Eq.~(\ref{eqsfr}) with $R=0$. Then the ratio between the SFR in  fresh gas,
${\rm SFR_{gf}}$, and the total SFR is given by, 
\begin{equation} 
{\rm{SFR_{gf}}\over{SFR}}\simeq {{1-R+w}\over{1+w}}. 
\label{cosmosfr} 
\end{equation}
The ratio decreases with increasing returned fraction and with decreasing
outflows, but it is never small. 
Even for large returned fractions $R=0.5$ 
\citep[e.g.,][]{2005A&G....46d..12E,2011ApJ...734...48L,2012ApJ...746..108T}
and moderate outflows $w=0.2$
\citep[e.g.,][; see also Sect.~\ref{winds}]{2012ApJ...757...54Z},  
${\rm SFR_{gf}/SFR}$ is as large as 58\,\%. 
If $w \gg 1$ then Eq.~(\ref{cosmosfr})  yields
${\rm SFR_{gf}\simeq SFR}$, which implies that when outflows 
are very important most of the SF during a 
gas consumption time is due to gas fallen in over that time. 

The mass in metals can be expressed using a differential equation
similar to Eq.~(\ref{1steq}) \citep[e.g.,][]{1990MNRAS.246..678E}.
It can be formally solved as indicated above, and the stationary-state solution 
gives a metallicity, 
\begin{equation}
Z\simeq Z_i +y\,(1-R)/(1-R+w).
\label{my_metal}
\end{equation}  
The symbol $y$ stands for the stellar yield, i.e., 
the mass of new metals eventually ejected per unit mass locked into stars.  
$Z_i$ represents the metallicity of the 
infall gas, which is usually much smaller than the metallicity of the disk gas.
In this case Eq.~(\ref{my_metal}) points out that the gas metallicity is independent 
of  mass inflow rate, SFR or mass of gas. $Z$ just depends on stellar
physics ($R$ and  $y$) and on the outflows through the mass loading factor $w$. 
This holds true even when  ${\rm \dot{M}_{in}}$, SFR and the stellar mass ${\rm M_\star}$
vary with time, provided they do it with a timescale longer than $\tau_g$.
Equation~(\ref{my_metal}) implies  that in a stationary state, the rate at which metal mass 
in the ISM is locked up into stars equals the rate at which new metal mass is ejected into 
the ISM by stars. This is how the  metal mass fraction in the ISM can stay constant
in the stationary state. 
The ISM does not necessarily have such an equilibrium 
because the disk gas mass and star formation rate fluctuate in time 
as a result of accretion fluctuations. 
Z in Eq.~(\ref{my_metal}) depends on the mass loading factor $w$ to 
compensate for the metal lost in outflows -- increasing $w$ is equivalent to 
decreasing the effectiveness of SF to produce metals.
It is as if $y$ were smaller which forces the 
equilibrium metallicity to be smaller as well. 
These results were originally found  
by \citet[][]{1972NPhS..236....7L} in the context  of the G~dwarf  
problem discussed in Sect.~\ref{star_accretion}.

\paragraph{Deviations from the stationary state.}
The stationary state described above represents
a useful idealization that has to be abandoned  
when the infall rate changes in a short time scale.
This is expected to happen quite often in view of the
clumpy and stochastic nature of the infall accretion
(Sect.~\ref{physical_picture}). Thus deviations from the 
stationary state are needed to understand the observed 
relationship between stellar mass, metallicity and 
SFR treated in Sect.~\ref{fmmr}. In addition, a young 
universe could have a relatively long gas consumption time, and then the SFR 
cannot keep up with the accretion rate. Observations of this imbalance 
were suggested by \citet{2012ApJ...754...25R} who found at $z>2$ an 
accretion rate from the KS relation that is 2 to 3 times larger than the 
SFR during galaxy build-up.

\paragraph{Average mass infall rate.} 
Cosmological accretion drives star formation. It is primarily
dark matter accretion carrying along baryons --  
intermittent but  
with a marked global trend. 
The average mass accretion rate into haloes of mass 
${\rm M_{halo}}$ 
at redshift $z$ may be approximated by a formula in
\citeauthor{2013MNRAS.435..999D}~(\citeyear{2013MNRAS.435..999D}; 
see also
\citeauthor{2013ApJ...772..119L}~\citeyear{2013ApJ...772..119L} and 
\citeauthor{2012MNRAS.421...98D}~\citeyear{2012MNRAS.421...98D}),
\begin{equation}
{{d{\rm M_{halo}}}\over{dt}}\approx 30\,{\rm M_{halo,12}}^{1.14}(1+z)^{2.5}\;{\rm M_\odot\;yr^{-1}},
\label{eqmdot}
\end{equation}
where ${\rm  M_{halo,12} =M_{halo}/10^{12}\,M_\odot}$. 
Approximating ${\rm M}^{1.14}$ by ${\rm M}$, and integrating equation (\ref{eqmdot}) over
time for an Einstein-deSitter Universe (which relates time and redshift), the
mass and accretion rate become \citep{2013MNRAS.435..999D}
\begin{equation}
{\rm M_{halo,12}\approx M_{halo,12,0}\,e^{-0.79(z-z_0)}}~,
\end{equation}
\begin{equation}
{{d{\rm M_{halo}}}\over{dt}}\approx 30\,{\rm M_{halo,12,0}}\,e^{-0.79(z-z_0)}(1+z)^{2.5}.
\label{halo}
\end{equation}
The fiducial mass ${\rm M_{halo,12,0}}$ corresponds to the mass at 
$z=z_0$, which would be today's mass if $z_0=0$.
The baryonic accretion rate is approximately the dark matter accretion rate 
multiplied by the cosmological baryonic mass fraction, $f_b$.  
The gas accretion rate depends on the faction of these 
baryons that get to the star-forming part in the center of the halo, 
$f_{\rm gal}$, so that, 
\begin{equation}
{\rm \dot{M}_{in}}=f_{\rm b}\, f_{\rm gal}\, {{d{\rm M_{halo}}}\over{dt}}.
\label{in_gas_como}
\end{equation}
For the Milky Way~(MW) today with
${\rm M_{halo,12}}=1$, $z=0$, $f_{b}=0.16$, and  $f_{\rm gal}\sim 1$ 
\citep[e.g.,][]{2014MNRAS.438..262P}, Eqs.~(\ref{eqmdot}) and (\ref{in_gas_como}) 
render an infall rate ${\rm \dot{M}_{in}}\sim5\;{\rm M_\odot yr^{-1}}$.

%
\subsection{Expected properties of accreted gas}\label{physical_picture}
Generally speaking,
gas accretion from the cosmic web occurs in two  modes, 
hot and cold, both taking place simultaneously in all galaxies
and over cosmic time. However, the mass of the DM halo determines 
which one is favored, so that the less massive galaxies prefer 
the cold-flow mode (Fig.~\ref{dekelf10}). 
The hot mode is produced by the gas being shock heated when 
entering the DM halo. It reaches a temperature 
${\rm T}> 10^5$K and a density between a hundred and a thousand
times the mean density of the universe 
\citep[][]{2003MNRAS.345..349B,2005MNRAS.363....2K}.
Gas at this temperature and density exhibits a
cooling time of several Gyr \citep[e.g.,][]{2006MNRAS.368....2D,2009MNRAS.393...99W},
which is long enough for the gas to be spread over the halo by the time
it settles down onto the galaxy disk.  The high temperature gas is too tenuous to 
be observed in X-ray, and is fully ionized.  
Cold-flow gas, however, rapidly ends up in the center of the potential 
well in a  free-fall timescale,
where it meets, strikes and heats the galaxy. Depending on the
flux and clumpiness of the gas stream, it may destroy a pre-existing
disk and create a new one, or just contribute to the growth
of the disk \citep[e.g.,][]{2009Natur.457..451D}.
The cold-flows are partly ionized with 
$10^4 \leq {\rm T}\le 10^5\,$K. 
The sources of ionization are mainly 
thermal collisions with electrons, and
photoionization due to individual galaxies 
and the UV background \citep[e.g.,][]{2011MNRAS.418.1796F}. 

\begin{figure}
\includegraphics[width=\linewidth]{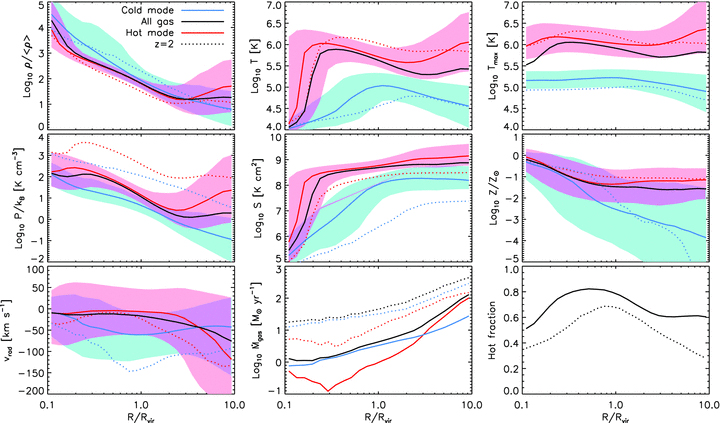}
\caption{
Properties of gas at $z=0$ in MW-type haloes. The physical parameters
are presented as a function of the distance to the halo center for all the gas (black), for hot gas (red), and 
for cold gas (blue).  Properties at $z=2$ have been included for comparison
(see the inset in the top left panel). 
Shaded regions correspond to $\pm 1\sigma$ scatter around the median. 
From the top-left to the bottom-right, the different panels show the mass-weighted 
median gas overdensity, 
temperature, maximum past temperature, pressure, entropy, metallicity, radial peculiar velocity, 
mean accretion rate and mean mass fraction of hot-mode gas, respectively. 
The radial profiles at $z= 2$ follow the same trends as at $z= 0$, 
although at $z=2$ the pressure, 
the infall velocity and the accretion rates are higher, the entropy is lower, 
and the hot mode accounts for a smaller fraction of the halo mass.
The distances are normalized to the virial radius ${\rm R_{vir}}$, and 
the density $\rho$
to the mean density of the universe $\langle\rho\rangle$. 
Taken from \citet{2012MNRAS.423.2991V}.
}
\label{conditions}       
\end{figure}

A summary of numerical model predictions is shown in 
Fig.~\ref{conditions} \citep[from][]{2012MNRAS.423.2991V}.
It describes the physical conditions of the gas being accreted by 
MW-size haloes    ($10^{11.5} < {\rm M_{halo}} < 10^{12.5}$\,M$_\odot$) 
at z= 0. 
The cold gas is defined to be the component that never reaches a temperature 
above   $10^{5.5}\,$K (Fig.~\ref{conditions}; top right panel). This cold component comes directly 
from the intergalactic medium (IGM) and is more metal poor than the hot component
 (Fig.~\ref{conditions}; middle right panel) which is enriched by galactic outflows. 
The gas falls in pulled by gravity  so that the characteristic velocities are 
similar to the Keplerian velocities in the outer 
parts of  disks.  The typical inflow velocity
for the cold-flow goes from 50 to 100\,\kms\ at $z=0$ although it 
increases with redshift for the same halo mass
 (see the bottom-left panel of Fig.~\ref{conditions}). 
The hot component has essentially no infall speed except near the outer parts. 
Distances to the center of the potential are normalized to the virial radius 
${\rm R_{vir}}$ defined as the radius than contains a 
density greater than $\sim$170 times the mean density of the universe. 
${\rm R_{vir}}$ is typically very large compared to the optical half-light 
radius of the resulting galaxies
\citep[$\sim 70$ times according to ][]{2013ApJ...764L..31K}.
Note that the average densities of both the hot and cold components are similar to within a factor 
of $\sim$5 and that both increase towards the center.  Still, the accretion rate is dominated by the 
cold component (lower middle panel) because it has the greater infall speed. 
The temperature of the hot component stays high until it finally reaches the disk, 
at which point it drops down to the temperature of the cold component. 
%

The dimensions of the cold streams determine how the gas interacts with 
the galaxy at the center of  the gravitational well, whether it is focused onto a small 
portion of the galaxy or affects the galaxy disk as a whole. Numerical simulations 
predict the comic gas streams to be very broad -- larger than the virial radius 
of a dwarf galaxy.  They are 0.1\,Mpc across at $z=2$ in the MW-like galaxy 
simulations by  \citet[][]{2010MNRAS.402.1536S}, and
the width increases with time, so that they are even broader
in the nearby universe. 
However, these broad streams develop  small-scale structure 
(called {\em substreams}), which is poorly captured at the 
resolution of the present simulations. These substreams
may eventually be small enough to become disk-galaxy size or 
smaller, and so they may identifiable 
through inhomogeneities produced in the disks \citep[e.g.,][]{2010MNRAS.404.2151C}.

The infall rates in the center of the potential well decrease with time, 
from a few tens of \smyr\ at $z=2$ to a few \smyr\ at $z=0$ 
(e.g., \citeauthor{2005MNRAS.363....2K}~\citeyear{2005MNRAS.363....2K}
\citeauthor{2012MNRAS.423.2991V}~\citeyear{2012MNRAS.423.2991V},
and also Eqs.~[\ref{halo}] and [\ref{in_gas_como}]).
These accretion rates are characteristic of MW haloes; higher masses have 
larger infall rates and a much steeper decrease with time. 
The accretion rate is bursty since the 
cold-flow gas falling in is clumpy rather than 
smooth \citep[e.g., Fig. 1 in ][]{2009ApJ...694..396B}.
As we addressed in more detail
in Sect.~\ref{stationary_state}, galaxies tend to be in a slowly evolving 
equilibrium state, with the predicted infall rates comparable to the SFRs. 
At every radius  accretion is balanced by SF, winds and 
radial transport of gas through the disk \citep{2014MNRAS.438.1552F}.

The mass in baryons in galaxy haloes can be several times larger than the 
stellar mass of the galaxy.  In massive haloes, most of halo gas is in the 
hot phase (Fig.~\ref{conditions}, bottom right panel), 
ionized at high temperatures. The cold gas is not uniformly distributed 
but covers only a small halo volume  \citep[$< 10$\,\%,][]{2011MNRAS.418.1796F}.
On average, the majority of stars present in any mass halo at any $z$ were formed 
from the gas accreted in the cold mode, although the hot mode contributes 
typically over 10\,\%\ for ${\rm M_{halo} \geq 10^{11}\, M_\odot}$
\citep{2011MNRAS.414.2458V}.
At redshift zero the fraction of stellar disk mass formed from shock-heated 
gas is between 5\,\%\ and 15\,\%\ when 
${\rm M_{halo}}$ goes from ${\rm 10^{11}\, M_\odot}$ to ${\rm 10^{12}\, M_\odot}$
\citep{2009ApJ...694..396B}.

%
\subsection{The importance of galactic winds}\label{winds}

Mass outflows, also known as galactic winds, have important 
effects on the evolution of galaxies, their haloes and the IGM. 
Together with mass accretion, 
they regulate the SFR of the galaxies, so that  the larger the winds 
the smaller the gas available to form stars and, therefore, the
smaller the SFR. Winds also transport the metals produced 
by stars out of the galaxy disks,  and so reduce the galaxy metallicity 
and pollute the circum-galactic medium (CGM)  and the IGM.
All of these features are present in the numerical simulations, but the
underlying cause can be appreciated using the analytic model
worked out in Sect.~\ref{stationary_state}, where the strength of the 
wind is parameterized in terms of the mass loading factor $w$. 
At a given accretion rate, 
when $w$ increases, the SFR decreases (Eq.~[\ref{eqsfr}]),
the mass of gas decreases (Eq.~[\ref{mykslaw}]), and the 
metallicity decreases (Eq.~[\ref{my_metal}]). 

Winds explain several central properties of galaxies. 
Without winds, there would be too many metals in the disks 
given the nucleosynthetic 
yield and the observed gas-to-star mass ratio (Sect.~\ref{fmmr}).  
Moreover, galaxies of different masses tend to have different metallicities,
therefore, if Eq.~(\ref{my_metal}) has to describe this property, the mass loading factor
$w$ must vary with the stellar mass.  Such dependence is to 
be expected, since galaxies of lower mass have shallower 
gravitational potentials and so they lose gas more easily 
(see Sect.~\ref{scaling_law}). 
Winds also affect the metallicity of the IGM, whose expected properties are 
put forward in Sect.~\ref{prediction_obs} and whose
observational properties are described in Sect.~\ref{hiiregions}.
Evidence for metal enrichment of the CGM is analyzed in 
Sect.~\ref{neutral_gas}.
  
The impact of winds on metallicity is so important 
because $w$ can be large, far in excess of unity. If $w \gg 1$,  then most of the 
accreted gas is not used to form stars but dragged 
along with the mass ejected in galactic winds
(${\rm SFR \ll \dot{M}_{in}}$ when $w\gg 1$ in Eq.~[\ref{eqsfr}]).
In order to reproduce the 
observed mass-metallicity relationship, \citet{2012MNRAS.421...98D} 
and \citet{2011MNRAS.417.2962P}
need $w$ to change from 1 to 6 when ${\rm M_\star}$ varies from 
$10^{11}$ to $10^9\,{\rm M_\odot}$. The same range is also
found in numerical simulations 
\citep[$w$ goes from 1 to 10 for ${\rm M_\star}$ between
$2\cdot 10^{11}$ and $10^9\,{\rm M_\odot}$; ][]{2012ApJ...760...50S}.  
Thus low mass galaxies are extremely wasteful using the 
cosmological gas,  most of which returns to the IGM 
without being processed through the stellar machinery.
In addition,  the nature of the wind of low and high mass 
galaxies may not be the same. The HI mass function of 
galaxy haloes depends on the wind model; it was matched
in a simulation by \cite{2013MNRAS.434.2645D} using momentum-driven 
outflows for large galaxies and energy driven outflows for small galaxies.
Low velocity winds from dwarf
galaxies can reach the intergalactic medium, while even high velocity winds from
massive galaxies may not as a result of the higher halo density that blocks
them \citep{2012MNRAS.420..829O}.

Winds do not seem to prevent cold accretion. 
\citet{2011MNRAS.417.2982F} showed from simulations that a
factor of 2 increase in the mass loading factor or a factor of 2 increase in the
wind speed affect most sensitively the accretion rate at the virial radius
without much effect on the cold gas accretion rate to the center. When both the
loading factor and speed increased by this factor, the cold accretion rate did
decrease for low halo masses. \citet{2011MNRAS.414.3671P} also found 
that galactic winds do not strongly affect the cold gas accretion rate.

The cycle of metals ejected by winds is complex, often involving re-accretion. 
 \cite{2012ApJ...760...50S}  model a galaxy at $z=3$ with a 
$2.4\cdot 10^{11}  {\rm M_\odot}$ halo and a ${\rm 2.1\cdot 10^{10}  M_\odot}$ stellar disk 
(like the present day MW disk), and include in the model satellites and nearby dwarfs. 
They find that 60\,\% of the accreted metals within 3 ${\rm R_{vir}}$ of the 
galaxy come from SNe in the host itself, 28\,\% come from satellite progenitors, 
and 12\,\% from nearby dwarfs.  Most metals beyond 2 ${\rm R_{vir}}$ are released very 
early, at $z$ from 5 to 8. An early release of highly ionized metals was
found by \cite{2013arXiv1309.5951F} also. Low ionization metals like MgII tend
to re-accrete within 2 Gyr, while high ionization metals take much longer.

Strong winds appear to be important for metallicity gradients in disks.
\cite{2013MNRAS.434.1531F} model analytically disk evolution with accretion,
winds, and star formation, assuming a radial accretion in the disk plane with a
velocity proportional to radius. They find that the largest influence on the
metallicity gradient is the re-accreted and enriched halo gas, rather than
metals mixed with the disk gas. To match observations of weak metallicity
gradients in galaxies with stellar masses of ${\rm \sim10^{10}\;M_\odot}$, 
they need 80\,\% of the metals produced by SNe to go into the halo where 
they can return to the disk, mostly in the outer parts. \cite{2012ApJ...750..122B} 
also suggest that metal-enriched outflows and re-accretion to the disk 
might explain their observation of metallicities in outer disks of two galaxies 
that are too large for the low rates of star formation there.

SNe affect accretion in another way, by providing ejecta of cool 
gas into the lower regions of the halo onto which the hot halo gas above it can 
condense. \cite{2010MNRAS.404.1464M} show in simulations that gas in a galactic 
fountain \citep{2008MNRAS.388..573M} mixes with hot halo gas, and if the 
pressure and metallicity of the halo are high enough, the mixture cools to 
become HI. 
The total amount of captured  halo gas can be about 20\,\% of the fountain gas for each event. 
%
\cite{2011A&A...525A.134M} and \cite{2012MNRAS.419.1107M} model 
the fountain and resulting halo accretion in detail, reproducing the 
intermediate velocity HI clouds above the MW and obtaining an accretion 
rate comparable to the star formation rate of $\sim 2\,$\smyr. 
This may solve the problem posed by the low inflow rate in MW HVCs discussed in Sect.~\ref{neutral_gas}.
\cite{2012MNRAS.419.1107M} suggest that the fountain ejection speed is around 70 
km s$^{-1}$. Absorption lines from the warm (e.g., SiIV) component of the 
MW, and from about half of the hot (OVI) component are also reproduced by this 
model \citep{2013MNRAS.433.1634M}, with these lines coming from fountain-mixed 
halo gas within a few kpc of the disk. \cite{2013MNRAS.434.1849H} simulate the 
formation of cold clumps in fountain filaments.

%
\subsection{Expected observational properties}\label{prediction_obs}

\paragraph{Metallicity.}
One of the fingerprints of the cosmic web gas is its low metallicity 
(between $10^{-2}$ and $10^{-3}\,Z_\odot$ at low redshift;
see Fig.~\ref{conditions}, central row, right panel), with a very large 
scatter indicating that  the metals are not uniformly spread  
\citep{2011MNRAS.418.1796F}.
Actually the question arises as to why metallicity is not 
zero if the gas is of cosmological origin. The answer is, partly,
a small contamination  from the first population III stars. 
They provide a floor of about $10^{-4}\,Z_\odot$, above which these very massive 
stars are no longer produced \citep[e.g.,][]{2004ARA&A..42...79B},
but the threshold may be even lower if the metals generated
by these stars do not reach the IGM
\citep{2013ApJ...772..106M}. However, the bulk of the metals is due to outflows
from  galaxies that come along with the gas streams
\citep[][]{2012ApJ...760...50S,2013arXiv1306.5766B}.
Accordingly, the metallicity of the web  increases with decreasing redshift, 
reaching the level of $10^{-2}\,Z_\odot$  at $z=0$ 
\citep[see Fig.~\ref{conditions}, and also][]{2012MNRAS.420..829O}.

\paragraph{Gas inflow.}
Inflow is the other distinct feature of accretion. 
Unfortunately,  observations do not easily distinguish inflows, 
outflows or other proper motions of the gas around galaxies. 
Doppler shifts alone do not  suffice since an inflow in the 
foreground and an outflow in the background produce the same Doppler shift. 
Additional physical constraints must be invoked.   For example,
the gas expelled from the galaxy tends to escape
in the direction perpendicular to  the galaxy plane, therefore,
motions in the plane of the galaxy favor the inflow 
interpretation \citep[e.g.,][]{2013Sci...341...50B,2014ASPC..480..211C}. 
As we discuss above,
cosmic web inflows are expected to be metal-poor so 
metal-rich moving gas is commonly interpreted as
outflow rather than inflow \citep[][]{2013ApJ...770..138L,2013ApJ...766...57B}.
A third possibility is using the relative velocity as a discriminant. 
The inflows cannot acquire a velocity largely exceeding the 
Keplerian velocities (Sect.~\ref{physical_picture}) and therefore  large 
velocities must be associated with outflows. 
The only direct way of assessing  accretion, however,
is by observing the gas in absorption in the starlight spectrum
emitted by the galaxy \citep[e.g.,][]{2013A&A...553A..16L} -- since
the gas is in between galaxy and observer, redshift implies infall.

\begin{figure}
\includegraphics[width=\linewidth]{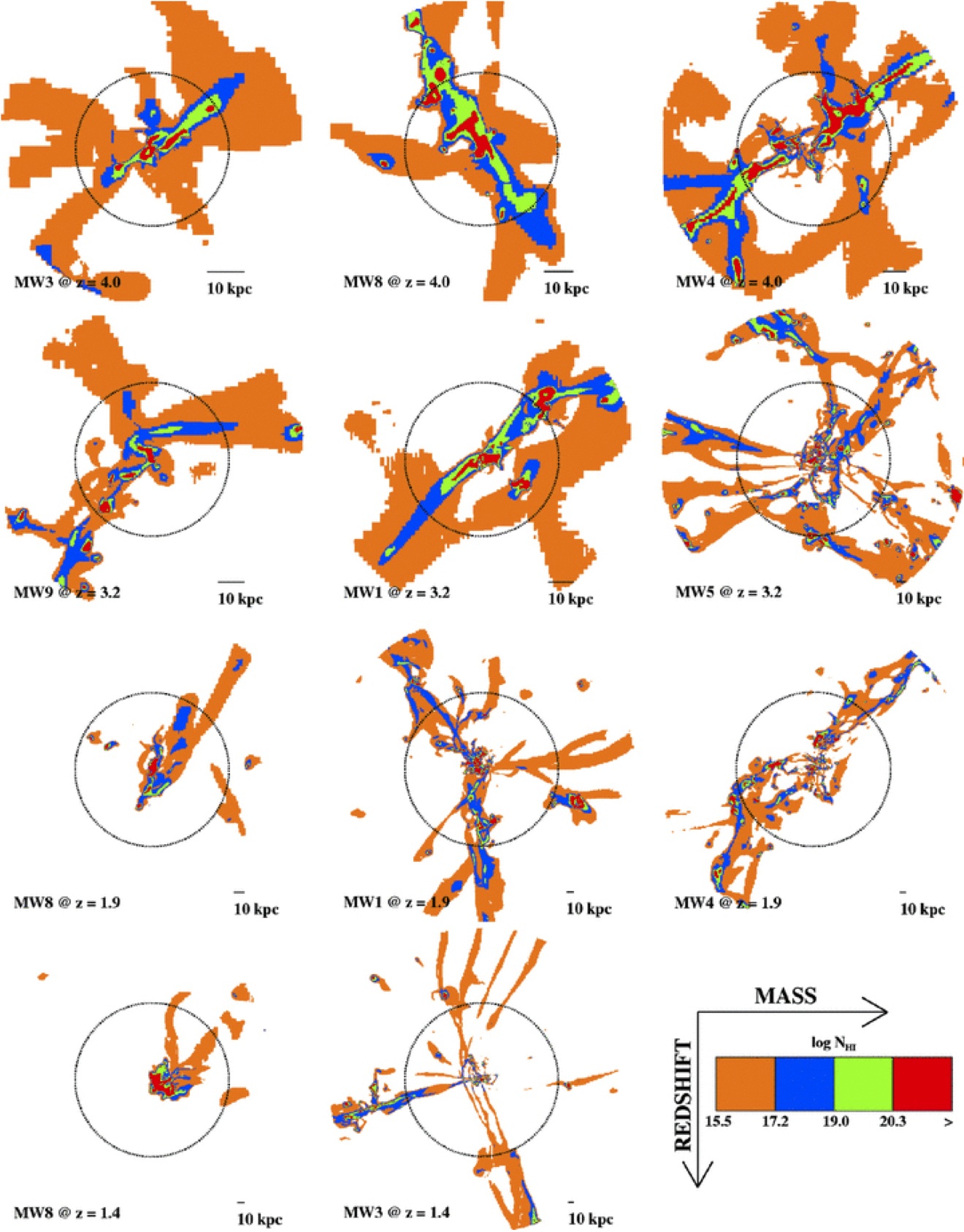}
\caption{
HI column densities in the cosmic web 
for a number of model galaxies in cosmological simulations as
worked out by \citet{2011MNRAS.418.1796F}.   
Color codes four intervals of column density including LLSs (in blue) to DLAs (in red).
Redshift is decreasing from top to bottom, and virial mass is increasing 
from left to right.  The dotted circles mark the virial radius. 
The cold streams are very patchy, with pockets of neutral gas immersed in
an ionized medium.
}
\label{himaps_simul}
\end{figure}

\paragraph{Ly$\alpha$ forest.}
One of the most consistent predictions of numerical simulations
has been the Ly$\alpha$ forest of low column density 
absorption lines to quasi stellar objects (QSO) produced by the cosmic web 
\citep{1992A&A...266....1B, 1996ApJ...457L..51H, 2001ApJ...559..507S}.
Simulations reproduce the distribution of column densities, equivalent widths
and power spectra \citep[for recent work, see][]{2014arXiv1401.6472B}.
\citet{2011MNRAS.418.1796F} work out the HI column density 
expected from the web considering the ionization balance to be 
controlled by thermal collisions  with electrons, by photoionization
due to individual sources and the UV background, and by dust shielding 
from ionizing photons. They predict column
densities of neutral hydrogen ${\rm N_{HI}}$ between $10^{17}$ and $10^{21}$\,cm$^{-2}$
(Fig.~\ref{himaps_simul}).
The expected properties  coincide with those observed in  
Lyman limit systems  (LLS; ${\rm N_{HI}}>10^{17}$\,cm$^{-2}$) and
 damped Lyman-$\alpha$ absorbers (DLA; ${\rm N_{HI}}>10^{20}$\,cm$^{-2}$),
including inflows of  100\,\kms\ and metallicities 
between $10^{-2.5}$ and $10^{-1}\,Z_\odot$ for $z$ between
5 and 0 \citep[e.g.,][]{2010MNRAS.402.1911T,2014ApJ...782L..29R}.
The closer to the center of the gravitational potential the denser the gas, and 
the lower the redshift the more diffuse. In simulations also
including radiative transfer, \cite{2012MNRAS.423.2991V} and
\cite{2012MNRAS.421.2809V} find that at $z=2$-3, most HI absorbers above
$10^{17}$ cm$^{-2}$ reside in haloes, those above $10^{21}$ cm$^{-2}$ are in
disks, and cold flow accretion should appear as an infalling component with 
low metallicity. \cite{2011MNRAS.418.1796F} suggest that cold streams
from galaxies in the mass range $10^{10}-10^{12}\;{\rm M_\odot}$ account for more than
30\,\% of the Ly$\alpha$ absorption in QSO spectra \citep[see also][]{2011MNRAS.412L.118F}.

%
\paragraph{Imaging the gas streams.}
Cold-flow streams contain partly ionized  gas undergoing continuous 
recombination and should produce an HI emission line  spectrum.  
\citet{2010MNRAS.407..613G} work out the Ly$\alpha$	flux to be 
expected from the cold streams that fed galaxies at high redshift. 
UV background excitation, collisions with free 
electrons, and dust attenuation are included.  They find that the emission is mainly 
driven by collisions since the filaments are thick enough to be shielded from the 
background. The peak surface brightness is 
$2\cdot 10^{-17}$\,erg\,cm$^{-2}$\,s$^{-1}$\,arcsec$^{-2}$ 
for MW-like haloes at $z=2.5$. 
Assuming that this flux is observed
through a narrow bandpass of 50\,\AA , it corresponds to 25.5 AB mag per
square~arcsec. The structures have sizes from 50 to 100 kpc,  
luminosities between $10^{43}$ and $10^{44}$\,erg\,s$^{-1}$, 
and are identified by some authors as the observed Ly$\alpha$ blobs
\citep[LABs;][see Sect.~\ref{cosmic_emission}]{2000ApJ...532..170S,
2001ApJ...562..605F,2009ApJ...696.1164O,2011MNRAS.413L..33L}. 
Simulations by  \citet{2012MNRAS.423..344R} also suggest that cold 
stream accretion can account for the Ly$\alpha$  emission.
The cosmic web gas also emits through fluorescence of Ly$\alpha$ 
photons produced by nearby sources \citep{2012MNRAS.425.1992C}.
This fluorescence has been recently observed 
as we discuss in Sect.~\ref{cosmic_emission}.

%
\subsection{Accretion and disk growth}\label{disks}

Accretion tends to favor the outer parts of galaxies with both stars from minor
mergers \citep[][]{2011MNRAS.416.2802F, 2012MNRAS.425..641L}
and gas from cold streams \citep{2010MNRAS.408..783R, 2012ApJ...745...66M}.
Surveys of extended UV disks around both early and late type galaxies 
actually suggest  on-going gas accretion and star formation in the outer
parts \citep{2011ApJ...733...74L, 2012ApJ...745...34M}. 
These ideas are consistent with cosmological numerical simulations, where disk
galaxies acquire their spin together with their mass through cold gas accretion.  

\cite{2011MNRAS.418.2493P} carry out simulations where the disk angular momentum 
comes from filaments and increases over time,  building up disks from the inside out.
\cite{2012MNRAS.427.3320C}, \cite{2013arXiv1310.3801L}, and
\cite{2014arXiv1402.1165D} show from cosmological simulations that low mass
galaxies, which tend to be disklike, spin in a direction that is aligned to
their accretion filaments, but high-mass spheroidal galaxies spin perpendicular
to their filaments. The difference arises because the disks accrete gas directly
from their filament and inherit its angular momentum, while the spheroids grow
from the mixture of other galaxies coming in along the filament (i.e,
grow from mergers). 
\cite{2011ApJ...738...39S} found that cold streams in haloes contain
higher angular momentum than dark matter.
The reason for 
this is that the total angular momentum of the dark matter adds positive and 
negative contributions from the mixing of many sub-haloes, with a net value that 
is small. In contrast, gas cannot pass through itself but dissipates relative motions 
and coheres into a smooth flow with the net angular momentum appearing as a systematic rotation. 

So far no numerical simulation details the interaction 
between a cold-flow stream and a galaxy disk in the local
universe.  \citet{2012MNRAS.420.3490C} model the 
kind of clumpy disks  observed at high redshift  
\citep{2005ApJ...631...85E,2007ApJ...658..763E,2011ApJ...733..101G}. 
The advent  of the gas stream changes the morphology of the disk so 
that it becomes more lopsided. Stars are formed immediately 
after gas arrival at high rate ($\sim 10^2$\,\smyr), but the 
metallicity of the resulting HII regions is not well modeled because part 
of the gas forming stars was in the disk already, 
and also because these simulations assume an 
instantaneous recycling of the SN ejecta. 
The size of the ensuing star-forming regions has to do with the gas 
accretion  rate and the turbulence, rather than with the cross section
of the gas stream that induces the process.
The turbulence in numerical disks may \citep[][]{2012MNRAS.425..788G} 
or may not \citep[][]{2013MNRAS.432.2639H} be maintained
by cold-flow accretion events.
Turbulence enhances mixing  processes on the scale of the disk thickness, 
so the mixing timescale for the ISM in a small radial annulus of a disk galaxy 
is expected  to be short -- of the order of a rotational period or shorter
\citep{1996AJ....111.1641T,2002ApJ...581.1047D,2005A&G....46d..12E,2012ApJ...758...48Y}.

\section{Accretion inferred from gas observations}\label{gas_accretion}

\subsection{Neutral gas observations}\label{neutral_gas}

\paragraph{Pools of neutral gas.}
The presence of pools of neutral gas around almost all galaxies is 
well known from the early days of radio astronomy 
\citep[e.g.,][]{1951Natur.168..357M,
1984ARA&A..22..445H,
1990ARA&A..28..215D}.
The improvement in sensitivity and the diversification of observational
techniques have reinforced the original view. 
Roughly speaking, the HI mass scales with the SFR of the galaxies
(KS law; see Sect.~\ref{stationary_state}) and since the SFR varies 
systematically  with the Hubble type and with ${\rm M_\star}$, gas rich 
systems tend to have late morphological types and low stellar masses.  
The gas-to-star mass ratio  ${\rm M_g/M_\star}$ varies from 0.01 to 10 when 
${\rm \log(M_\star/M_\odot)}$ goes from 11.5 to 8,
with the same range of variation from red early type galaxies
to blue late types \citep{2004ApJ...611L..89K,
2004A&A...425..849P,
2013MNRAS.434.2645D}.
The scatter of the relationships with stellar mass and Hubble type
is significant, so that neither the mass nor the Hubble type fully determine the 
gas fraction of a galaxy \citep{2004ApJ...611L..89K,2013ApJ...777...42K,2011MNRAS.413.1373W}.

\begin{figure}
\includegraphics[width=\linewidth]{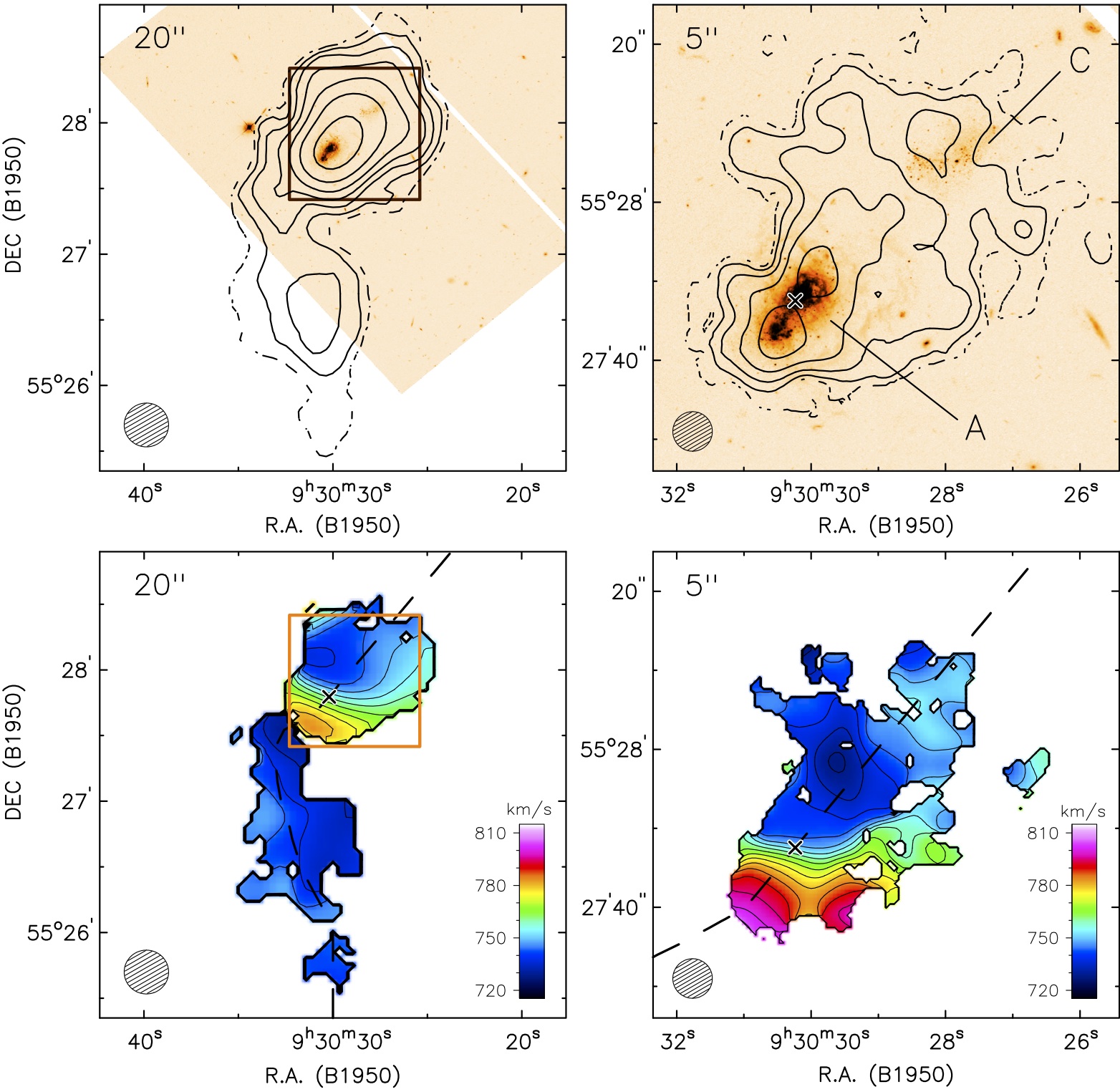}
\caption{
Figure from \citet{2012A&A...537A..72L} corresponding to IZw18. Top: 
integrated HI  maps at a resolution of 20\,arcsec (left) and 5\,arcsec (right), 
superimposed on a HST image
\citep[from][]{2007ApJ...667L.151A}.
The box in the left panel shows the area covered by the right panel. 
In the map at 20\,arcsec resolution, ${\rm N_{HI}}$
contours are at 0.25 (dashed), 0.5, 1, 2, 4, 8, 16  
$\times 10^{20}$\,cm$^{-2}$. 
In the map at 5\,arcsec resolution, contours are at 3 (dashed), 6, 12, 24, 48  
$\times 10^{20}$\,cm$^{-2}$. 
The circles show the beam size. A and C refer to the components labeled by 
\citet{1989AJ.....97.1591D}.
Bottom: HI velocity fields corresponding to the HI maps on top. The velocity scale is 
color coded as given in the insets.
}
\label{izw18}
\end{figure}

\citet{2008A&ARv..15..189S} summarize radio observations of HI cloud complexes, 
tails and  filaments in and around local galaxies, suggesting ongoing minor mergers and 
recent arrival of external gas. The HI image  of the dwarf galaxy  IZw18 in Fig.~\ref{izw18} is 
particularly  illustrative \citep[from][]{2012A&A...537A..72L}. It shows large amounts of 
HI gas around the galaxy 
(${\rm M_g/M_\star}\sim 5$) with a plume of different velocity
directed towards the galaxy  where a large  starburst goes on  at present
(Fig.~\ref{izw18}, left panels). 
Although not as spectacular as in IZw18, elongated extra-planar features are quite 
common in disk galaxies (e.g., NGC925, \citeauthor{1998AJ....115..975P}~\citeyear{1998AJ....115..975P}, 
and the many other examples given in the review by
\citeauthor{2008A&ARv..15..189S}~\citeyear{2008A&ARv..15..189S}). 
Extra-planar gas is also found  in our own galaxy, where the existence of gas 
clouds falling in is  long known 
\citep[][]{
1966BAN....18..421O,
1999ApJ...514..818B,
2004ASSL..312.....V,
2007ApJ...670L.113W,
2008ApJ...672..298W}. 
Some of these high velocity clouds (HVC) are part of the so-called {\em galactic fountain}, where SN 
ejecta return to the galaxy plane after cooling down \citep[e.g.,][]{1996AJ....111.1641T,2008A&A...484..743S}. 
However, HVCs with low metallicities are likely of extragalactic origin
\citep[e.g.,][]{2013ApJ...777...19H}.
The existence of HVCs agree qualitatively with theoretical expectations, however, 
they  pose an observational problem common to many other galaxies. 
Their mass infall rate is estimated to be $\sim$\,0.1-0.2\,\smyr\ which is one order of 
magnitude too small  to maintain the MW SFR of $\sim$\,1-2\,\smyr\ 
\citep[][]{2008A&ARv..15..189S,2014IAUS..298..228F}, 
and thus inconsistent
with the MW being in a stationary state sustained by HVCs (Sect.~\ref{stationary_state}). 
This imbalance between observed SFR and observed gas accretion rate is  common  but
not universal \citep[counter examples are given
by, e.g.,][]{2013Sci...341...50B,2014A&A...561A..28S}. 
Moreover,
\citet{2012ApJ...750..165R} has mapped HVCs in local galaxies.
 His HI observations and models 
indicate that the gas accretion rate density of $z\sim0$ galaxies is 0.022 
${\rm M_\odot\, yr^{-1}\, Mpc^{-3}}$, 
which is comparable to the local universe SFR density 
(Sect.~\ref{redshift} and Fig.~\ref{bouwens04}).
He notes that most local accretion may be warm 
rather than cool gas,  although cool infalling HVCs undoubtedly are important 
components of local galaxy star formation
\citep[e.g.,][]{2014AJ....147...48P}.
A massive (${\rm 10^{10}\,M_\odot}$) hot ($10^6$\,K) gas is indeed 
detected over a large region (100\,kpc) around the MW \citep{2012ApJ...756L...8G}.
%
%
Part of the gas supply needed to balance the MW SFR may be due to 
ionized HVCs. \citet{2011Sci...334..955L} measure the ionized HVC mass 
to be $\sim 10^8\;{\rm M}_\odot$ and  estimate an infall time of $80$-130\,Myr, 
which gives an accretion rate of 0.8-1.4$\;{\rm M}_\odot$\,yr$^{-1}$, large enough to
sustain the SFR in the MW. The gas supply may also come from the Magellanic stream.  
The MW is surrounded by a long stream of neutral and ionized gas that is leading and trailing 
the Large and Small Magellanic Clouds \citep{1972ApJ...173L.119W,1974ApJ...190..291M}.
This gas presumably came from the Magellanic clouds during their interaction with 
each other \citep{2012ApJ...750...36D} and is now experiencing a drag force from its 
motion in the hot MW halo gas \citep{2005MNRAS.363..509M}.  
Taking into account the neutral and ionized gas, the  Magellanic stream
contains $2\cdot 10^9\;{\rm M}_\odot$ at a distance of 55~kpc 
\citep{2014arXiv1404.5514F}. If the gas in the Magellanic system survives to reach
 the MW disk over its inflow time of $\sim 0.5$-1.5\,Gyr,  it will represent an 
average inflow rate of 4-7\,\smyr\ potentially raising the current MW SFR. 
Multiple signs of an evaporative interaction with the 
hot halo indicate that the stream may not survive its journey to the disk 
fully intact. It will break apart by hydrodynamic 
instabilities  and evaporate,  contributing the hot halo mass 
\citep{2007ApJ...670L.109B,2014arXiv1404.5514F}
Another big piece of evidence for gas reservoirs in galaxies 
comes  from  Ly$\alpha$ absorption produced by gas around 
galaxies that happen to be near QSO lines-of-sight.
In a systematic search for such systems,
\citet{2013ApJ...777...59T}  find absorption in all star-forming 
galaxies, with a detection rate as large as 75\,\% for the early types
and 100\% for late types.
The sample contains all Hubble types in the  near universe ($z < 0.35$) 
with intermediate masses $9.5 \leq \log{{\rm (M_\star/M_\odot)}} \leq 11.5$, and 
the lines-of-sight sample impact parameters $d < 150$\,kpc. 
The gas is associated with the galaxies (with velocities smaller than the escape 
velocity) and it is found to be in a cold phase, with temperatures well 
below the halo virial temperatures ($< 10^5$\,K).
This reservoir of gas is often metal poor, indicating 
that it was not generated and ejected by the galaxies 
but it is of cosmic origin (Sect.~\ref{prediction_obs}).
\citet{2011ApJ...743L..34K} examine Mg II absorption kinematics in halo gas at impact 
parameters from 12-90 kpc  along 11 QSO sightlines towards 13 L$_\star$ 
galaxies\footnote{L$_\star$ is the characteristic luminosity that separates 
low and high luminosities in the galaxy luminosity function by 
\citet{1976ApJ...203..297S}. It turns out to be roughly comparable 
in luminosity to the MW.} 
at $z<0.15$ and
find that thick disk rotating halo models are not sufficient to explain the broad lines. 
They suggest the gas is infalling along filaments and streams.  
In a blind survey, \citet{2013ApJ...779L..17B} found C~IV 
absorption that appears to be associated 
with either of two dwarf galaxies, and concluded that the most likely origin for 
the lines is cold gas accreting onto the
 $z\sim0.003$ dwarf galaxy.
The same kind of cold gas is also present in galaxies of lower
mass and to even larger impact parameter.  \citet{2014arXiv1402.3602J} 
search for galaxy-QSO close pairs suitable for finding gas absorption 
systems. They find 111 independent pairs among 
nearby $(z\sim 0.02)$ low mass  galaxies   
${\rm (\log [M_\star/M_\odot] = 9.0\pm 0.9)}$, with impact parameters
between 30 and 500\,kpc (median $250$\,kpc). 
Abundant hydrogen gas is found beyond 
the dark matter halo radius ${\rm R_{vir}}$  and all through $d\sim 500$\,kpc, 
with a mean covering fraction of $\sim 50$\%. No heavy  elements are 
detected at $d> 0.5\,{\rm R_{vir}}$, though. The authors estimate that the metallicity of the 
inner halo is  0.1\,$Z_\odot$ (Chen 2014, private communication), which sets 
an upper limit to the metallicity of the IGM at 
distances larger than ${\rm R_{vir}}$. 
The above results present just a few cases among the extensive list 
of evidence  for neutral gas reservoirs next to galaxies drawn from 
absorption on QSO spectra -- more will be given in Sect.~\ref{redshift}.

Within the cold-flow accretion scenario, gravitationally bound 
HI structures devoid from stars are to be expected. HI filaments of this kind 
have been discovered in blind HI surveys. For example, 
\citet[][]{2011A&A...533A.122P} analyze data from the Parkes All Sky Survey
to find ten faint extra-galactic filaments that can 
correspond to extended haloes, tidal remnants, or potentially diffuse filaments 
tracing the neutral fraction of the cosmic web. \citet{2011A&A...527A..90P}
map the galaxy filament connecting the Virgo cluster with the local group, 
finding 20 new  detections of neutral hydrogen with no obvious sign
of stellar emission. Dark galaxies are also found in the HI  ALFALFA survey
\citep[][]{2007NCimB.122.1109H,2014AAS...22324651M}. 
They represent some 3\% of the extragalactic sources \citep{2008IAUS..244...83H}.
Other examples exist in the recent literature \citep[e.g.,][]{2013A&A...555L...7O}.
So far it is unclear whether the small number of HI-only sources 
agrees with cosmological numerical simulations since
the predictions are not specific enough.  

Environment influences gas content and recycling. An HI ALFALFA survey of 365 dwarf galaxies in the 
Virgo Cluster revealed 12 red early-type dwarfs with 
the same HI per unit mass as star-forming dwarfs \citep{2012AJ....144...87H}. The most likely 
interpretation is that the early type dwarfs were once stripped in the cluster are recently re-accreting gas.
 In a VLA HI study of counter-rotating disks, three early-type barred galaxies appear to be accreting cold gas. 
Several of the HI features are cold gas blobs, possibly gas-rich dwarfs 
\citep{2012MNRAS.422.1083C}.  \citet{2011AJ....141..204K} selected voids out to $d<$100 Mpc based on 
SDSS galaxy data in order to search for HI in regions of greatest galaxy underdensities. Among 15 galaxies, 
14 were detected with HI gas masses from $3.5\cdot 10^8$\,-\,$3.8\cdot 10^9 {\rm M_\odot}$. While some of the galaxies are 
interacting or have companions, others appear isolated and undisturbed, with flocculent spiral or chain morphologies. 
They note that one void galaxy is a polar ring galaxy in a thin wall between voids, and may be slowly accreting gas from
 the cosmic web, as also for another ring galaxy in a void \citep{2013MNRAS.434.3310S}.
Two dwarf galaxies with extended HI disks have unusual kinematics that may be explained by ongoing cold flow accretion. 
One, the only known dwarf galaxy in the Local Void (KK 246), has an extended HI disk of ${\rm \sim10^8 M_\odot}$ and an 
unusually high mass-to-light ratio of 90 \citep{2011AJ....141..204K}.

In order to characterize faint diffuse gas in local galaxies, \citet{2011A&A...526A.118H} have initiated the 
HALOGAS (Westerbork Hydrogen Accretion in LOcal GAlaxieS) survey. Their deep HI survey is designed to search for
 extraplanar gas in edge-on and inclined nearby galaxies and determine the distribution and gas properties. The pilot 
survey of 4 galaxies tentatively suggests that galaxies  with higher star formation energy form gas haloes from outflow, and that 
extraplanar gas is associated with a high star formation rate per unit area \citep{2012ApJ...760...37Z}. The full survey should also 
be a useful characterization of halo gas in local systems.

%
\paragraph{Morphological distortions.} Very often the HI maps present 
non-axisymmetric distortions showing that the gas is not contained in a disk 
or a spheroid \citep[e.g.,][]{1980MNRAS.193..313B} and so, suggesting the gas 
distribution to be transient.  For example, spirals are known to have extended 
warped HI distributions, a pattern that cannot be sustained for long.  
Among other explanations,
gas infall has been proposed as the  origin of warped disks 
\citep[e.g.,][]{2002A&A...386..169L,2005A&A...438..507B,2013MNRAS.434.2069K}. 
Figure~\ref{izw18} 
shows an extreme case of distorted HI  morphology. Such extreme  distortion is 
actually common among galaxies with  large specific SFR in the local universe --  
blue compact dwarf (BCD) galaxies often show such complex HI topology 
\citep[][]{2006MNRAS.372..853E,
2010MNRAS.403..295E,
2012A&A...544A.145L,
2013ApJ...779L..15N}. 
The survey LITTLE THINGS finds BCDs with kinematically separate 
components,  streamers extending far beyond the optical size,
kinematic HI axes offset from the optical axes, clouds associated with 
recent starbursts, and so on  
\citep{2012AJ....144..152J,2013AJ....146...42A}.
Many other examples of HI maps showing galaxies with plumes and 
filaments  can be found in the literature 
\citep[e.g.,][]{2012MNRAS.419.1051L}.
Morphological distortions suggestive of gas accretion are also common 
in optical images and are discussed in Sect.~\ref{star_accretion}. 

A particularly telling deviation from axi-symmetry has been found
recently  by \citet{2012ApJ...760L...7K}. 
They report a bimodality in the azimuthal angle 
distribution of low ionization gas around galaxies as traced by MgII 
absorption along QSO lines-of-sight.  The circum-galactic gas prefers to lie near the 
projected galaxy major and minor axes. The bimodality is clear in 
blue star-forming galaxies whereas red passive galaxies exhibit 
an excess of absorption along their major axis. These results suggest 
the bimodality to be driven by gas accretion along the galaxy major 
axis and gas outflows along the minor axis (Sect.~\ref{prediction_obs}).
There is no other clear alternative. For this 
interpretation to be correct, the inflow has to be highly anisotropic, 
concentrated in the plane of the galaxy. This fact favors cold-flow
accretion with respect to the isotropic gas inflow to be expected
from hot coronal gas cooling down (Sect.~\ref{physical_picture}).   

The lack of axi-symmetry of the gas properties in galaxy haloes is corroborated 
by \citet{2013ApJ...773...16Z} in their study of CaII~H\&K absorption lines
produced by one million foreground galaxies at $z \sim 0.1$ in $10^5$ 
SDSS QSO spectra. For edge-on galaxies the absorption is more concentrated 
along the minor axis. This is consistent with the idea that bipolar outflows 
induced by star formation produce the metallic gas in the haloes.
Actually, the inflows along the galaxy plane help to channel the 
outflows along the poles \citep[][]{1997ApJ...478..134T,1998MNRAS.293..299T}.

%
\paragraph{Metallicity.} Having low metallicity can be regarded as 
the main fingerprint of cosmic web gas (Sect.~\ref{prediction_obs}) and, 
indeed, the  neutral gas around galaxies is often very metal poor. 
Gaseous systems observed in absorption along QSO lines-of-sight are
easier to detect if they are metallic, simply because the absorption
features are intrinsically stronger.
In order to avoid this bias, 
\citet{2013ApJ...770..138L}  use an HI selected sample of some 
30 $L_\star$ galaxies with redshifts up to one.
They study the metallicity of  LLS  
(${\rm 10^{16} < N_{HI} < 10^{18.5}\,cm^{-2}}$) 
with impact parameters $d$ between 10 and 150\,kpc.
Interestingly, they find a bimodal metallicity distribution with  
metal poor and metal rich branches peaking at 0.025 and 0.5\,$Z_\odot$, 
respectively. 
Both branches have a nearly equal number of absorbers.
These two populations fit in well with the scenario depicted in Sect.~\ref{theory},
with coexisting metal-rich gas outflows from the galaxies 
and cosmological gas inflows. We note that the cosmic web metallicity at $z\sim 0$
predicted by numerical models  is very close to the metal-poor branch observed  by 
\citeauthor{2013ApJ...770..138L}
In a complementary work,  \citet{2014arXiv1402.3602J} find 
that the absorbing gas next to the galaxies has metallic lines, but 
only hydrogen absorptions show up at distances beyond the
virial radius. This pattern is corroborated  by other 
studies \citep[e.g.,][]{2013ApJ...779...87C}.

Extremely metal poor
(XMP) galaxies provide another piece of evidence for 
large amounts of cosmic gas around local star-forming
galaxies. These galaxies have a number of properties consistent 
with disks being assembled by accretion of  gas
\citep{2012ApJ...750...95E,2013ApJ...767...74S,2014ApJ...783...45S}. 
In particular, they are  enshrouded by large amounts of neutral 
gas, so that ${\rm M_g/M_\star}$ is typically as 
large as 20 \citep{2013A&A...558A..18F}.  The ratio is so large
that normal stellar evolution cannot produce enough metals
for their HI gas to have the same metallicity as that observed in the HII
regions of the galaxies, which are already extremely metal poor
because XMP galaxies are selected to have
$Z< 0.1\, Z_\odot$. 
Then the HI gas around XMPs
must have a metallicity 
$Z\ll 0.1\, Z_\odot$ and so
of the order of a few hundredths of the solar 
value,  which is the level of metallicity expected for the cosmic web 
in the local universe  (Sect.~\ref{prediction_obs}).

The CGM has also been studied in
absorption in the stellar spectra produced by  the host galaxy. Then
one can unambiguously  measure whether the gas goes in or out (see Sect.~\ref{prediction_obs}). 
\citet{2013A&A...553A..16L} make this observation with IZw18 (Fig.~\ref{izw18}).
The HI gas pool around it has a metal abundance lower by a factor of two as 
compared to the HII regions, and it may even present pockets of  gas 
with metallicity essentially null. The HII metallicity  samples the gas in the 
disk, which is already extremely low in this particular galaxy. 
Since the observed HI gas metallicity is at the level expected for the
cosmic web (Sect.~\ref{prediction_obs}), and it is lower than the
metallicity in the disk, the HI is likely cosmic gas being accreted.
\citet{2009A&A...494..915L} find an even larger discrepancy between
the HI and HII metallicities of Pox~36, where the HI metallicity is around
0.03\,$Z_\odot$.  

%
\paragraph{Gas consumption timescales.}
As we point out above, the stellar mass is  a poor predictor of the gas-to-stellar
mass ratio ${\rm M_g/M_\star}$, 
however, \citet{2013ApJ...777...42K} find that ${\rm M_g/M_\star}$ correlates well with 
 the stellar mass in young stars formed only during the last  Gyr,  ${\rm M_\star^{new}}$. 
Actually, ${\rm M_g/M_\star}$ was found to be close to ${\rm M_\star^{new}/M_\star}$, which
implies routine refueling of star-forming galaxies on Gyr timescales, i.e., 
implies a continuous gas accretion to maintain SF. The study is based on thousands 
of objects so it portrays a general property of galaxies.
This short time for exhausting the gas is very consistent with the HI consumption timescales
inferred from the KS~law ($\taug$ in Eq.~$[$\ref{kslaw}$]$), of the order of 0.5-2\,Gyr 
(Sect.~\ref{stationary_state}). The fact that galaxies have been forming stars along the 
Hubble time \citep[e.g.,][]{2012ApJ...756..163S}, but 
they consume their gas in only one Gyr, is one of the best 
arguments in favor of the need for an external gas supply to keep up with
the star formation \citep[e.g.,][]{1983ApJ...272...54K,2012MNRAS.426.2166F}.

%

\subsection{Ionized gas observations}\label{hiiregions}

The emission line spectra produced by HII regions surrounding the
star-forming regions provide a direct means of measuring the physical conditions
of the gas forming stars and, in particular, its metallicity that 
represents the prime diagnostic for cosmic gas accretion (Sect.~\ref{prediction_obs}). 
There are a number of compelling observations of ongoing gas accretion 
based on such HII-based metallicity measurements.

\paragraph{Metallicity inhomogeneities and inverted gradients.} 
The secular evolution of disk galaxies produces a regular pattern with the metallicity 
decreasing inside out, i.e., having a negative gradient with galactocentric distance 
 \citep{1988MNRAS.235..633V,1995MNRAS.272..241E,1998AJ....116.2805V,
2007A&A...470..843M,2014A&A...563A..49S}.  
The timescale for gas mixing in a disk inside an annulus is 
fairly short, on the order of a rotational period or a few 
hundred Myr (Sect.~\ref{disks}).
Deviations from negative metallicity gradients
are usually attributed to the recent arrival
of cosmic gas that feeds the star formation. If the 
gas is accreted through the cold-flow mode (Sect.~\ref{theory}),
it is expected to reach the disks in clumps often 
forming stars already
\citep[e.g.,][]{2009Natur.457..451D,2010MNRAS.404.2151C,2012ApJ...745...11G}. 
Alternatively,  the external gas streams may fuel the disks with metal-poor gas,
so that  gas mass builds up developing
starbursts through internal gravitational instabilities  
\citep[e.g.,][]{1999ApJ...514...77N,2008ApJ...688...67E,2009ApJ...694L.158B}.   
In any case, the cold-flow accretion is bound to induce 
metal-poor starbursts. 

\begin{figure}
\includegraphics[width=1.\linewidth]{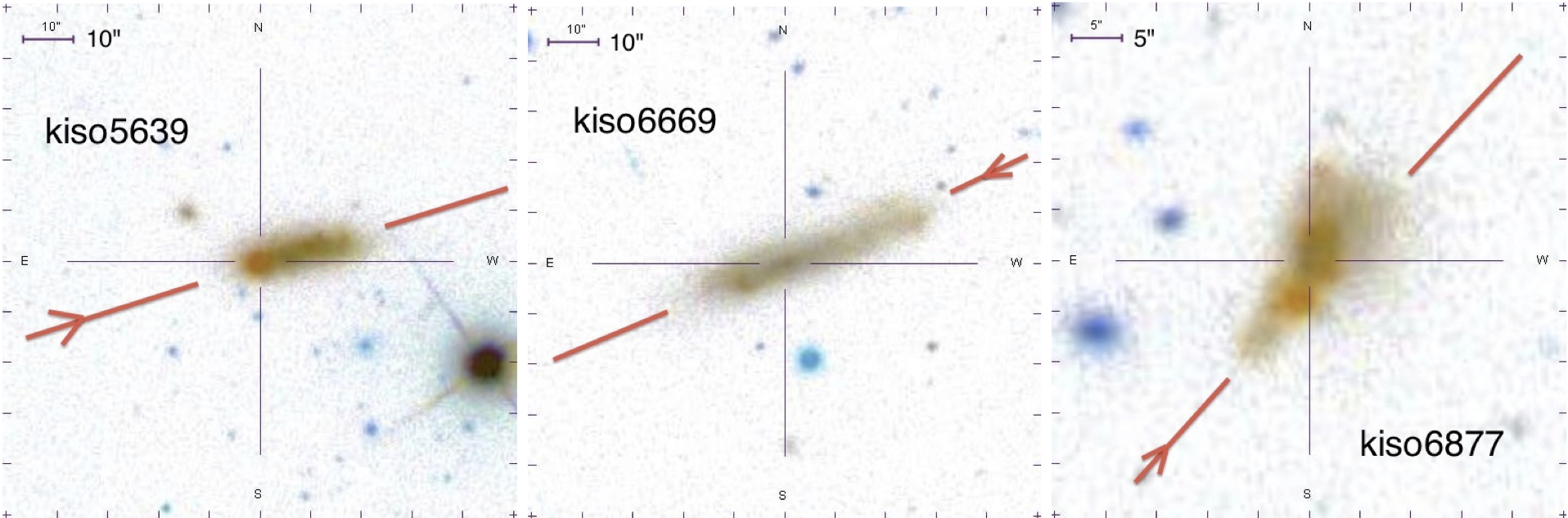}

\vspace{5mm}

\includegraphics[width=1.\linewidth]{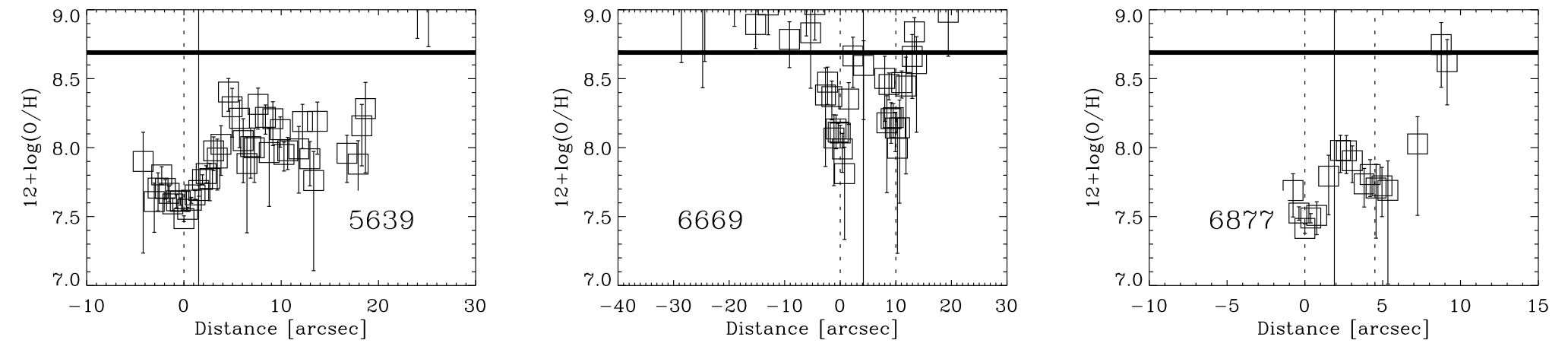}
\caption{
Top: 
Images of three tadpole galaxies characterized by having a bright peripheral clump on a faint tail.
The images have been taken from the SDSS database \citep{2009ApJS..182..543A},  
and are displayed with an inverted 
color palette  so that the background sky looks white, and the intrinsically blue galaxies 
appear  reddish. The red line shows the orientation of the spectrograph slit used to measure 
metallicity variations, and  the horizontal scales on 
the upper left corner of the  panels correspond to 5 or 10 arcsec as indicated.
Bottom:
Oxygen abundance variation across the galaxies on top. 
The vertical solid line represents the center of rotation, 
whereas the vertical dotted lines indicate the location of maxima in SFR. 
Note the existence of abundance variations, with the minima coinciding with the 
regions of largest  SFRs.
The thick horizontal solid line indicates the solar metallicity.
Adapted from \citet[][]{2013ApJ...767...74S}. 
}
\label{tadpole}
\end{figure}

Metallicity drops associated with intense starbursts may reflect cold-flow accretion. 
\citet{2010Natur.467..811C} were the first 
to identify this pattern in high redshift galaxies.
The same kind of metallicity drops appear in local tadpole
galaxies, where the bright star-forming head of the tadpole 
often has lower metallicity than the underlying galaxy (see Fig.~\ref{tadpole}).  
\citet{2013ApJ...767...74S} interpret this observation  as 
an episode of gas accretion onto the tadpole head. 
Localized  metallicity drops associated with 
star-forming regions have also been observed in other 
objects,  including  gamma ray burst host galaxies \citep{2011ApJ...739...23L,2014arXiv1404.0881T},
BCDs  \citep{2009A&A...503...61I,2010ApJ...715..656W,2014ApJ...783...45S}
and dwarf irregular galaxies \citep{2013ApJ...765...66H}. Variations of metallicity among HII 
regions  located at the same galactrocentric distance are not unusual
even in large nearby spirals \citep[][]{2012ApJ...750..122B,2013ApJ...766...17L};
some of these variations could come from localized accretion events.

Secular evolution produces negative metallicity gradients, but some spirals show 
{\em reverse} gradients, where the metallicity is lowest in the inner galactic regions.
Examples are given by \citet{2012A&A...539A..93Q}, who point out that
$15\,\%$ of their galaxies at $z=1.2$ show such reverse gradients, especially
those objects with the lowest metallicities. 
This observation may be due to averaging out metallicity drops in inner 
regions, or other artifacts caused by the limited angular resolution
\citep[e.g.][]{2013ApJ...767..106Y}. Alternatively, it may also be 
interpreted as produced by fast inflows within the disk giving gas to the
central regions. Metal poor gas deposited in the outskirts can be
transported outside-in by bar instabilities or some type of tidal 
interaction \citep{2008IAUS..245..151C,2014ASPC..480..211C,2012ApJ...747..105E}.
The same mechanism of gas transport is able to explain the presence 
of metal-poor gas in the narrow-line region of a nearby 
QSO found by \citet{2011A&A...535A..72H}. 
It may also account for the finding by \citet{2012ApJ...745...66M}
that 10\,\%\ of the galaxies with regular metallicities  exhibit a sharp 
downturn  in metallicity at the edge of the disk. Surprisingly, the magnitude of the 
outer drop is correlated with the fractional HI content of the galaxy. 
The recent stellar mass growth at the edge of the galaxies is apparently
due to the accretion or radial transport of  gas from beyond the stellar disk.

\paragraph{High metallicity of quiescent BCDs.} 
BCDs are high surface brightness targets relatively easy to detect. 
The luminosity of these galaxies is
dominated by one or several young starbursts. However, most if not all BCDs contain host galaxies
 with old stars too
\citep[e.g.,][]{1996A&AS..120..207P,2003ApJ...593..312C,2006ApJ...651..861C,
2007A&A...467..541A}.
The dominant starburst is so intense that it cannot be sustained for long, therefore, the 
BCDs have to be in a transient phase.  Consequently, there must be many local galaxies in the pre or 
post BCD phase, i.e., many quiescent BCDs (or, for short, QBCDs).
The BCD host galaxies should show up best outside of their starburst regions. 
Masking out these regions,  \citet{2007A&A...467..541A,2009A&A...501...75A}
were able to characterize their photometric properties. 
Using the typical host colors and magnitudes as proxies for 
QBCD properties, \citet{2008ApJ...685..194S}
searched the SDSS-DR6 archive for QBCD candidates. They turned out to be rather common: one 
out of three local dwarf galaxies is of this kind, and there are some thirty of them per BCD galaxy. 
Their main properties, including their luminosity functions, are consistent with the BCDs being QBCDs 
observed during a starburst phase in a duty cycle where the quiescent phase lasts 30 times 
longer than the active phase. 
This interpretation presents a difficulty, though: the gas-phase metallicity of the QBCDs is
 systematically higher 
than the metallicity of the BCDs. This cannot happen in a closed-box evolution, where the 
precursor galaxy always has 
lower metallicity than the follower, so that QBCDs could not be precursors of BCDs. 
The problem naturally 
disappears if almost every BCD phase is preceded by the advent of fresh metal-poor gas 
that triggers the star
 formation episode. Moreover, such gas infall triggering  explains 
why the stellar  metallicities of BCDs and QBCDs
agree, even though their gas-phase metallicities do not \citep{2009ApJ...698.1497S}.
The stars of BCDs and QBCDs are statistically the same because only a small fraction of 
galaxy stellar mass is  produced in each starburst. Their gas differs because BCDs have 
just rejuvenated their ISM.
Keeping in mind that  30\,\%\ of all local dwarfs are QBCDs,
gas-infall must be a common phenomenon.
These findings are consistent with the recent  results by \citet{2013ApJ...764...44Z}.

\paragraph{Metallicity threshold.}
Local galaxies with ionized gas metallicity smaller than $0.1\,Z_\odot$ 
are rare. They are usually called extremely metal poor (XMP).
The recent compilation by \citet[][]{2011ApJ...743...77M} rendered 
only 140 such objects, all of them with metallicity above $0.01\,Z_\odot$. 
The existence of this threshold has been known for a long time since 
the prototypical XMP galaxy IZw\,18 is close to the limit (some $0.03\,Z_\odot$),
 and its singularity was acknowledged more than 40 years ago \citep{1970ApJ...162L.155S}. 
Despite repeated efforts to find galaxies 
more metal poor \citep[e.g.,][]{1991A&AS...91..285T,1999ApJ...527..757I,2000A&ARv..10....1K}
the 1\,\%\ lower limit metallicity remains today
\citep[][]{2005ApJ...632..210I,2011EAS....48...95K}. 
Several explanations have been put forward to account for
this minimum metallicity: the self-enrichment of the HII region
used for measuring \citep{1986ApJ...300..496K},
the metal abundance of the proto-galactic cloud 
\citep{2011EAS....48...95K}, the metallicity threshold set by  
the ejecta from population III stars  \citep{1995ApJ...451L..49A,2005ApJS..161..240T}, 
technical difficulties for metallicity determinations below a 
threshold \citep{2008A&A...491..113P},
and others. 
None of them seem to be fully compelling.
However, the accretion scenario provides a natural 
explanation for this long-lasting problem. Numerical simulations 
predict the cosmic web gas to accumulate metals from 
the outflows of dwarf galaxies (Sect.~\ref{winds}). These contributions add up
along the Hubble time so that at redshift zero the cosmic web metallicity
has to be  at the few percent level (Sect.~\ref{prediction_obs}), which is 
precisely the observational threshold.
This is the metallicity to be expected if the SF in XMPs is driven by gas directly 
accreted from the cosmic web. 

\paragraph{Nitrogen and Oxygen in green-pea (GP) galaxies.}
GPs are star-forming galaxies which receive this name because of their compactness and green color in 
SDSS composite images \citep[][]{2009MNRAS.399.1191C}. The color is produced by an unusually large
[OIII]$\lambda$5007\,\AA\  emission line redshifted so as to contribute to the g-band color. 
They have some of the highest specific SFRs in the local Universe, able to double their stellar masses in a 
fraction of Gyr. GPs are low metallicity outliers of the mass metallicity relationship 
\citep{2010ApJ...715L.128A,2012ApJ...749..185A}.
Detailed analyses of their emission lines reveals complex kinematical structures with several components 
coexisting in only a few kpc \citep{2012ApJ...754L..22A}. Even though GPs have low oxygen metallicity, 
they present an overabundance of ${\rm N/O}$ which is typical of aging stellar populations. 
This puzzling observation is naturally explained if GPs have recently received a major flood of low metallicity gas 
\citep{2010ApJ...715L.128A,2012ApJ...749..185A}. Then the mixing with metal-poor gas reduces the metallicity
 (i.e., O/H), but the ratio between metal species (N/O) remains as in the original high metallicity ISM.
The motions of a galaxy in the N/O vs O/H plane due to gas infall have been  
modeled by  \citet{2005A&A...434..531K}, who point out that large excursions require 
the infall gas mass to be larger than the gas present in the galaxy, with the infall rate exceeding the SFR. 
We note that GPs are not special but just extreme cases in the continuous sequence of local 
star-forming galaxies \citep[e.g.,][]{2011ApJ...728..161I}.

\section{Accretion inferred from stellar observations}\label{star_accretion}

\begin{figure}
\includegraphics[width=0.6\linewidth]{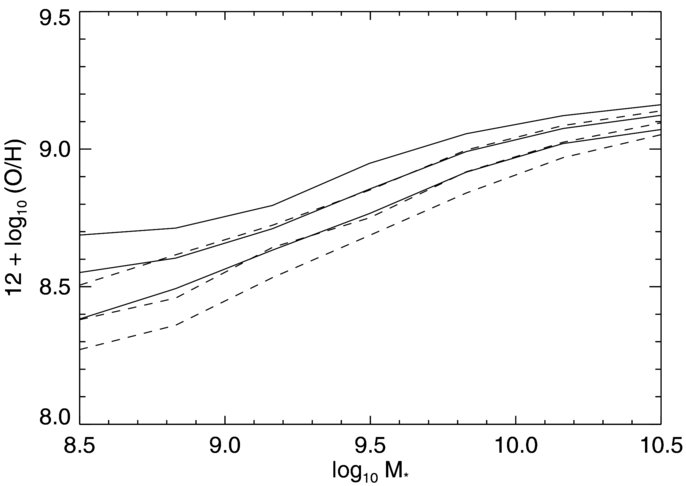}
\caption{
Stelar mass-metallicity relation for the most lopsided
(the dashed lines) and the least lopsided (the solid lines) 
galaxies in the sample of $2.5 \cdot 10^4$ nearby galaxies 
from SDSS studied by \citet{2009ApJ...691.1005R}. 
The three lines of each type correspond to the 25\,\% , 50\,\%, and 75\,\%\
percentiles. The most lopsided galaxies have a metallicity deficit of 
0.05-0.15 dex compared to the least lopsided galaxies of
the same ${\rm M_\star}$. The deficit at low mass is greater than at 
high mass. Oxygen abundance is used to parameterize  
metallicity, and ${\rm M_\star}$ is given in solar mass units.
}
\label{lopsideness}
\end{figure}

\paragraph{Morphology metallicity relationship.} 
There is a relationship between the morphology of the galaxy as inferred from broad-band imaging (thus
tracing stars) and the metallicity of the star-forming gas. It has several manifestations, the 
most conspicuous one being the association between cometary shape and extremely metal poor
galaxies \citep[][]{2008A&A...491..113P,2011ApJ...743...77M,2014arXiv1404.5170M}.
In a systematic search for XMP galaxies in SDSS-DR7, \citet{2011ApJ...743...77M} find that
75\,\%\ of them have either cometary shape or are formed by chained knots. Likewise,
from the comprehensive catalog of 140 known XMPs used by \citet{2013A&A...558A..18F}, 
80\,\%\ have cometary structure or two or multiple star-forming regions. 
For reference, only 0.2\,\%\ of the star-forming galaxies in the 
Kiso survey are cometary \citep{2012ApJ...750...95E}.
Even if surprising, XMPs seem to be the extreme case of a common relationship between morphology and 
metallicity followed by the bulk of the star-forming galaxies in the local universe. 
\citet{2009ApJ...691.1005R} measure lopsidedness in a sample of 
$2.5 \cdot 10^4$ nearby galaxies from SDSS. 
At a fixed mass, the more metal-poor galaxies are more  lopsided -- see Fig.~\ref{lopsideness}. 
Whatever process causes lopsidedness,  it is associated with low metallicity 
gas in the galaxies.  This non-trivial observational result is naturally
accommodated within the  gas accretion scenario 
\citep[e.g.,][and  Sect.~\ref{theory}]{2012MNRAS.420.3490C}. 
Extreme lopsidedness is produced by off-center large starbursts, fed by gas accretion either
directly or indirectly -- directly if the gas arrives to the disk ready to form stars  or indirectly if the gas
 is accumulated until disk instabilities trigger SF.  Thus low metallicity and lopsidedness 
come together naturally. The process has to be quite common to be responsible for the 
lopsidedness in the large dataset explored by \citeauthor{2009ApJ...691.1005R}

\begin{figure}
\includegraphics[width=\linewidth]{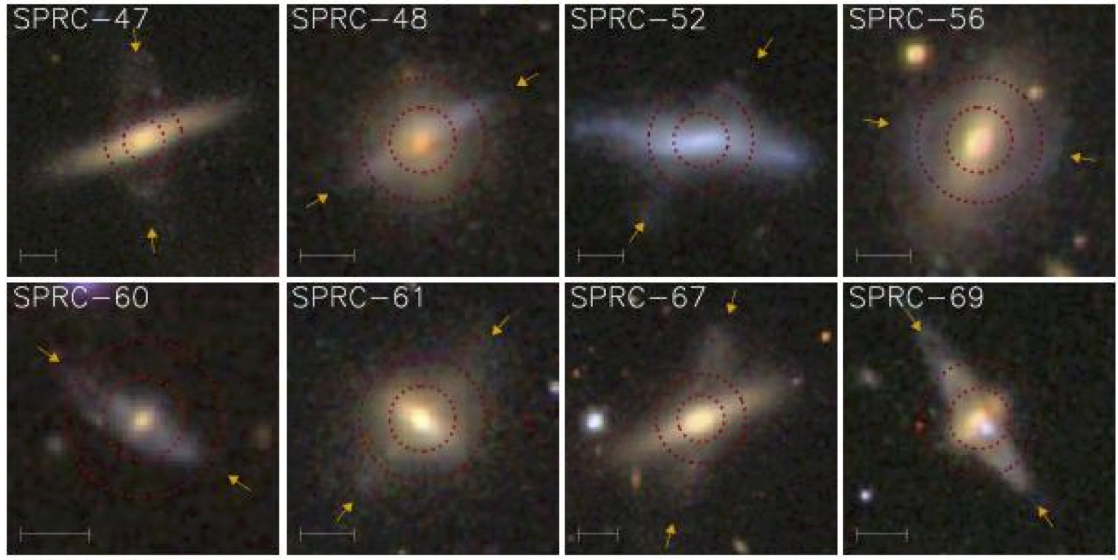}
\caption{
SDSS color images of polar ring  galaxies  from \citet[][]{2013A&A...554A..11C}. 
The scale bar is 10 arcsec long, and the rings are indicated with small arrows to guide
identification. The central object is usually an early-type gas poor galaxy, 
but the ring itself is gas rich. Some galaxies have giant HI polar disks with very weak 
stellar counterpart  \citep[e.g.,][]{2011MNRAS.418..244M}.
}
\label{polar_ring}
\end{figure}

\paragraph{Kinematical distortions.}
Galaxies are, to first order,  axi-symmetric systems.  However, they frequently 
show  kinematical  distortions with respect to an axi-symmetric velocity 
pattern. When the distortions are significant but  the underlying 
axisymmetric structure remains identifiable, the best explanation is 
often the recent accretion of a gas rich object or a gas stream
onto a pre-exiting galaxy. For example,
polar-ring galaxies are composed of a 
central component (usually an early-type disk galaxy) surrounded by an outer ring or 
disk made of gas, dust and stars,  which orbits nearly perpendicular to the 
plane of the central galaxy
(\citeauthor{1990AJ....100.1489W}~\citeyear{1990AJ....100.1489W}; 
see also  Fig.~\ref{polar_ring}).
The polar ring itself is gas rich, with a gas fraction corresponding to
$\sim$30\,\% of the baryonic mass, suggesting that polar ring galaxies  have just 
accreted a large amount of gas \citep{2013A&A...554A..11C} and new stars are
formed in situ. Polar rings can be formed through galaxy interaction and merging, 
but most likely through accretion from cosmic filaments 
\citep[e.g.,][]{2011MNRAS.418.1834F,2013A&A...560A..14P,2013MNRAS.435.1958B}. 
Kinematically distinct inner polar gaseous disks appear common in disk early-type galaxies; 
their juxtaposition with old stellar nuclei suggests that they may be remnants of gas accretion that occurred prior to
 the main gas accretion in the galaxy \citep{2014arXiv1401.3366S,2012MNRAS.423L..79C}. 
Polar ring metallicities of 0.1--0.4 $Z_\odot$,
 lower than the metallicities in the parent galaxies, are consistent with cold accretion 
\citep{2011A&A...531A..21S}.  

Often galaxy disks show a number of kinematically distinct components
like counter-rotating bulges \citep{1996ApJ...463L...9P},
counter-rotating disk stars \citep[][]{2007A&A...463..883V, 2011MNRAS.412L.113C}, 
and SF regions kinematically decoupled from the rest of
 the galaxy \citep{2013ApJ...767...74S,2014MNRAS.441..452K}. 
These features are expected from numerical simulations of minor 
mergers \citep[e.g.,][]{1990ApJ...361..381B, 2004ApJ...611...20I},
however they also arise  in 
hierarchical galaxy formation scenarios even in the absence of merging.
\citet{2014MNRAS.437.3596A} model a disk galaxy fed by two distinct filamentary 
structures with opposite spins. They produce counter rotating stars that
are not dragged along with the gas but mostly produced in situ.
Tidal streams or little tails are also kinematically distinct components,
and they turn out to be quite common in the local 
universe. At least 6\,\%\ of the local galaxies show distinct stream-like 
features, and a total of 19\,\% show some sort of faint tail 
\citep[e.g.,][]{2011A&A...536A..66M,2010AJ....140..962M}.
They are thought to be leftovers of tidally disrupted 
gas rich satellites on their way to reach the center of the 
global gravitational potential.
A study by \citet{1998AJ....116.1169M} 
indicated a correlation between optical and 
HI asymmetries in isolated galaxies, which could result from cold gas accretion.

The highly organized distribution of satellites  surrounding the MW
and M31 has been long noticed \citep[e.g.,][]{1983IAUS..100...89L}. 
The best explanation seems to be  that the dwarf  satellite galaxies fell into the 
gravitational potential along only one or two  filaments  \citep{2008MNRAS.385.1365L,2011MNRAS.416.1401A}.
Using a high-resolution dark matter simulation of the local group, \citet{2011MNRAS.411.1525L}
concluded that the satellites of both MW and M31 were accreted  anisotropically onto their haloes, 
entering the virial radius from a specific direction with respect to the large-scale structure.  
These conclusions are confirmed in other simulations of M31 dwarfs 
\citep{2013arXiv1307.2102G,2013arXiv1307.5044S}, 
MW dwarfs \citep[][]{2011ApJ...742..110N,2011MNRAS.413.3013L}, 
and in observations of dwarfs around NGC 3109 \citep{2013A&A...559L..11B}.

\paragraph{Star formation history of dwarf galaxies.}
The variation of the SFR with time is often very bursty in
late type galaxies, even if they are isolated. 
Episodes of large star formation are intertwined with long 
quiescent phases.  This result is inferred using fitting techniques on 
integrated galaxy spectra \citep[e.g.,][]{2007IAUS..241..461C},
color-magnitude diagrams of resolved stellar populations 
\citep{2010ApJ...721..297M}, and statistical arguments involving active 
and quiescent BCDs \citep[][]{2008ApJ...685..194S}. Convulsive SF histories
fit in well the cold-flow accretion scenario. Models
predict the process to be intermittent, so, if gas accretion drives 
star formation, then the SFR is expected to be bursty (Sect.~\ref{theory}). 
A number of  factors conspire to make the effect more clear in dwarfs. 
They are actively forming stars at present, so that the bursts are young 
and luminous, and they provide SF histories with good
time resolution. Outflows are far more
important in dwarfs, and they quench star formation
forcing long inter-burst periods. Forming massive galaxies
by cold-flows requires the contribution of many discrete 
accretion events, therefore, statistical fluctuations
in their arrival time tend to cancel, giving rise to a smooth 
SF history. The lower the galaxy mass the less effective the
statistical averaging, and  dwarfs tend to have a more spasmodic SFR.

Even though individual HII regions last for a few Myr 
\citep[e.g.,][]{1999A&A...349..765M,2004ApJ...603..503H},  
the SF episodes of a galaxy as a whole do not shut off within this 
timescale but they last for several hundred 
Myr \citep{1998AJ....116.1227D,2010ApJ...724...49M,2011ApJ...730...14H}. 
This relatively long duration of the burst is consistent with a single gas 
accretion event which suddenly increases the gas but which requires a 
much longer gas consumption timescale to transform it into stars  
($\taug\sim 1\,$Gyr; Sect.~\ref{stationary_state}).

%
\paragraph{G~dwarf problem.}
There is a long known deficit of sub-solar metallicity  G~dwarf stars in 
the solar neighborhood  \citep[][]{1962AJ.....67..486V,1963ApJ...137..758S,1975VA.....19..299L}. 
This so-called {\em G~dwarf problem} has been amply discussed
in the literature, with solutions ranging from changes in the initial mass function (IMF)
to inhomogeneous star formation \citep[][and references therein]{2008NewA...13..314C}.
Among them, a continuous metal-poor gas inflow feeding 
star formation seems to be the preferred one \citep[][]{1990MNRAS.246..678E,2005A&G....46d..12E}.
The argument dates back to \citet{1972NPhS..236....7L}, who pointed out  
that star-formation maintained by metal-poor gas accretion 
self-regulates to produce a constant gas phase metallicity close to the 
stellar yield, which implies close to the solar metallicity (Sect.~\ref{stationary_state}). 
Therefore the apparent deficit of sub-solar metallicity G~dwarfs is actually an 
excess of solar metallicity G~dwarf stars formed over time out of an
ISM always near equilibrium at approximately the solar metallicity. Interestingly, the 
G~dwarf problem is not exclusive to the MW but is known to
happen in other galaxies as well \citep[e.g.,][]{1996AJ....112..948W}.
The enhanced deuterium fraction in the Galaxy is also consistent with this 
scenario \citep[e.g.,][]{2011MNRAS.410.2540T}, and 
chemical evolution models cannot produce the abundances
observed in other stars or the abundance gradients without extensive
and continuous accretion of fresh metal-poor gas 
\citep[e.g.,][]{2009MNRAS.396..203S}.

\section{Scaling laws as evidence for metal-poor 
gas accretion}\label{scaling_law}

A number of observational properties characterizing large samples 
of  local star-forming galaxies can be explained if the
star-formation is driven by metal poor gas accretion. 
The gas infall explanation provides  a simple physical unifying 
mechanism, even though often it is not  the only explanation 
of each individual observation. 
The very existence of general laws or trends implies
that the underlying mechanism has to be something 
fundamental, since it  affects not just a few objects but the
bulk of the star-forming galaxies. 
Some of these general properties are discussed elsewhere 
in this work, and will not be repeated here; in particular
the short gas consumption time-scale compared with the 
age of the stars (Sect.~\ref{neutral_gas}),
the large metallicity of quiescent BCDs (Sect.~\ref{hiiregions}),
the metallicity-morphology relationship (Sect.~\ref{star_accretion}),
and the similarity between the mass in young stars and mass of gas 
(Sect.~\ref{hiiregions}).

\subsection{The stellar mass-metallicity-SFR
relationship}\label{fmmr}

\begin{figure}
\includegraphics[width=\linewidth]{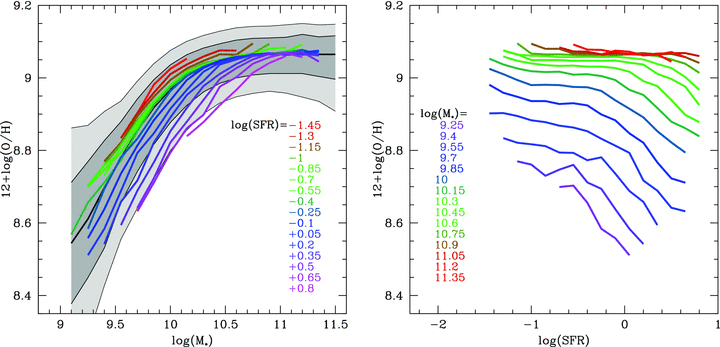}
\caption{Stellar mass-metallicity-SFR relationship,  or FMR, as observed 
by \citet{2010MNRAS.408.2115M}.
Left: oxygen abundance vs ${\rm M_\star}$ for galaxies with the same
SFR. SFRs are color coded as indicated in the inset.
Given a galaxy mass, galaxies of lower metallicity present 
higher SFR.
Right: 
oxygen abundance vs SFR for galaxies of the same mass.
The stellar mass is color coded according to the inset.
Given a SFR, galaxies of lower metallicity have lower mass.
The relationship tends to saturate at the high-mass end, where
all galaxies have the same metallicity independently of
the mass and SFR. ${\rm M_\star}$ is given in ${\rm M_\odot}$ and
SFR in \smyr .
}
\label{fmr10}
\end{figure}

Galaxies are known to follow a mass-metallicity relationship,
where the larger the stellar mass the higher the metallicity 
\citep[e.g.,][]{1989ApJ...347..875S,2004ApJ...613..898T,2005MNRAS.362...41G}.
The relationship presents a significant scatter that has been recently 
found to be associated with the present SFR in the galaxy 
\citep{2010MNRAS.408.2115M,
2010A&A...521L..53L,
2012MNRAS.422..215Y,
2013A&A...549A..25P,
2013ApJ...765..140A,
2013arXiv1310.4950Z}.
Specifically,  for galaxies with the same stellar mass, the 
metallicity decreases as the current SFR increases. This 
new relationship is often referred to as the fundamental 
metallicity relationship \citep[FMR,][]{2010MNRAS.408.2115M},
and is shown in Fig.~\ref{fmr10}.
The mass-metallicity relationship is commonly interpreted as
due to variations of the star-formation efficiency with galaxy 
mass, and/or to galaxy mass-dependent metal-rich outflows
\citep[e.g.,][]{2008IAUS..255..100L,2008ApJ...672L.107E}. 
The former implies that low mass galaxies produce less stars
for their gas, and so remain more metal poor, whereas the latter 
relies on the metal-rich SN ejecta to be preferentially lost to the 
IGM by low mass galaxies.
Neither of these two mechanisms, however, 
predict the observed  dependence of metallicity on the SFR. 
%
The observed anti-correlation between metallicity
and SFR  can be qualitatively understood if the star-formation is 
preferentially triggered and sustained by the inflow of metal-poor gas.
The arrival of new gas simultaneously drops the ISM metallicity and 
increases the  SFR. This idea was already advanced by 
\citet{2010MNRAS.408.2115M}, and it seems to work also
quantitatively as shown in the next paragraphs.

Several recent numerical and analytic works explain the
FMR within this framework of gas infall forcing SF. It is important
to acknowledge that such an explanation implies a time variable infall rate,
which moves the galaxy system out of the stationary-state.
Otherwise, in a gas consumption time-scale ($\sim$1\,Gyr), the
galaxies reach an equilibrium characterized by a metallicity independent
of the infall rate (see Eq.~[\ref{my_metal}]). The stationary-state 
metallicity depends only on parameters set by stellar physics and on the 
mass loading factor $w$ (i.e, the ratio between the outflow rate and 
the SFR; see Eq.~[\ref{mass_load}]).  Since the gas has to 
overcome the gravitational pull to escape, $w$ is expected to be a 
strongly  varying function of galaxy mass.
This dependence of $w$ on the depth of the gravitational well 
(as parameterized by ${\rm M_\star}$) 
seems to be the key ingredient giving rise to the observed 
mass-metallicity  relationship \citep[e.g.,][]{2002ApJ...581.1019G,
2008MNRAS.385.2181F,2011MNRAS.417.2962P}.  
Since $Z\propto {\rm M_\star}^{1/3}$ at low stellar mass, 
\citet{2012MNRAS.421...98D} conclude that  
$w\propto {\rm M_\star}^{-1/3}$,
as expected if the outflows are produced by momentum driven winds.

\citet{2012ApJ...750..142B} describe the FMR with an 
intrinsically bursty model which is always out of equilibrium. 
Galaxies receive metal-poor gas parcels of different masses
that are instantaneously mixed up with the existing ISM. 
This arrival dilutes the metallicity of the gas and 
triggers star formation, with the two parameters set by 
the received gas mass and thus correlated. 
The authors show that at fixed ${\rm M_\star}$, 
the observed relationship between metallicity and SFR 
follows the scaling to be expected from this model. 
The explanation works for most of the  galaxies with 
${\rm M_\star  \leq 2 \cdot  10^{10}\,M_\odot}$,
and many with masses above this threshold and large radii.

\citet{2013MNRAS.430.2891D} explain the FMR by forcing a 
secular deviation from the stationary state, so that given a  ${\rm M_\star}$,  those systems 
that happen to have larger gas masses are also those with higher SFR and lower $Z$. 
The reason why two galaxies with same ${\rm M_\star}$ have different gas 
masses can be pinned down to their different ages, so the youngest is the one 
with the largest gas mass. Both inflows and outflows are required 
to reproduce the observed relationship, but the outflows dominate by far.

\citet{2013ApJ...772..119L} work out a model galaxy 
whose main properties self-regulate due to the short
gas consumption time scale. Although more sophisticated, it 
is  very much in the spirit of the simple description given in 
Sect.~\ref{stationary_state}. The model galaxy is near the stationary state, 
but the gas supply available to form stars is allowed to change 
with time. This drives the system out of the stationary state and 
provides a dependence of the metallicity on the SFR and mass gas. 
The authors successfully fit the FMR by 
\citet[][in our Fig.~\ref{fmr10}]{2010MNRAS.408.2115M},  
allowing both $w$ and the gas consumption time scale to
vary with ${\rm M_\star}$. As in the case of \citet{2013MNRAS.430.2891D},
the deviations from the stationary state are set by the galaxy age.

\begin{figure}
\includegraphics[width=0.6\linewidth]{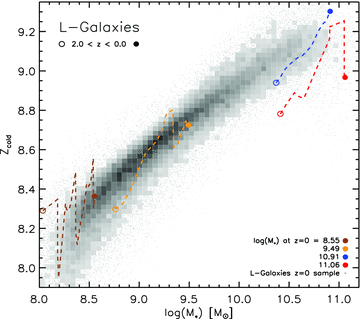}
\caption{
Evolution of  four model galaxies from $z=2$ (open circles)  to 0 (bullets) 
according to the semi-analytical models by \citet{2012MNRAS.422..215Y}. 
Galaxies of final stellar mass ${\rm \log(M_\star/M_\odot)}=8.55, 9.49, 10.91$ and 11.06  
are shown in brown, orange, blue and red, 
respectively. The full present day population is shown in grey for reference. 
Galaxies move up and to the right as they gather more mass and reduce their outflows. 
In addition, sudden metallicity drops are produced by the arrival of cosmic gas ready to form
stars.
}
\label{fmr_evolution}
\end{figure}

\citet{2012MNRAS.422..215Y} use the semi-analytical model 
{\em L-Galaxies} by \citet{2011MNRAS.413..101G} to explain the 
FMR. Thousands of individual galaxies are taken from purely 
dark-matter cosmological numerical simulations. 
Baryons that follow the dark-matter are added, implying that a 
non-stationary clumpy gas accretion drives the system.
Mass loading factors that depend on ${\rm M_\star}$ are 
included. The model reproduces the main observational trends, 
including an apparent turnover of the mass-metallicity  relationship  at very 
high stellar masses. Figure~\ref{fmr_evolution} shows the evolution
of four prototypical model galaxies in the $Z$~vs~$M_\star$ plane.
It illustrates how the FMR naturally emerges in this seemingly complex
picture. There is a global quasi-stationary drift up and to the right,
as the galaxies gather mass and so decrease their outflow losses. 
On top of this trend, galaxies move down and up  at an 
almost constant stellar mass. First they move down in response to the arrival of 
fresh gas, and then they quickly recover the stationary-state metallicity and move back up.
This  is  particularly clear in the galaxy of lowest mass in Fig.~\ref{fmr_evolution}
(${\rm \log[M_\star/M_\odot]}=8.55$,  brown line and symbols).
The stationary-state metallicity is given by the upper envelope of the 
$Z$ vs ${\rm M_\star}$ relationship.
Obviously, the downward excursions are accompanied by bursts of star 
formation, giving rise to the observed FMR when many galaxies at different 
phases of the cycle are represented in the $Z$ vs ${\rm M_\star}$ plane.

\citet{2013arXiv1311.1509F} present a simple model to understand the genesis of the FMR.
The scatter arises from the intrinsic scatter in the accretion rate, but may be substantially 
reduced depending on the timescale on which the accretion varies compared to the timescale 
on which the galaxy loses gas mass.  They show that observational constraints on the scatter 
can be translated into constraints on the galaxy-to-galaxy variation in the mass loading factor, 
and into the timescales and magnitude of stochastic accretion onto star-forming galaxies. 
They infer remarkably small scatter in the mass loading factor, $< 0.1$\,dex, 
and that the scatter in accretion rates is smaller than expected from numerical simulations.

The parameterization of the FMR by  \citet{2010MNRAS.408.2115M},
seems to hold until $z < 2$, and then it changes substantially at higher redshift 
\citep[e.g.,][]{2010MNRAS.408.2115M,
2012PASJ...64...60Y,
2014ApJ...780..122L,
2014A&A...563A..58T,
2014MNRAS.440.2300C}. 
Variations are expected if the relationship is caused by 
cosmic gas accretion since accretion rates and IGM metallicities 
change over time.  Thus the variation with redshift of the FMR  
provides a powerful diagnostic tool to evaluate the evolution of the accreted gas
along the history of the universe  
(\citeauthor{2013arXiv1311.1509F}~\citeyear{2013arXiv1311.1509F}; see also Sect.~\ref{conclusions}).

\paragraph{N/O is independent of SFR at a given ${\rm M_\star}$.}
Unlike the metallicity, the ratio between the observed
N and O does not seem to depend on SFR 
\citep[see][]{2013A&A...549A..25P,2013ApJ...765..140A}.
This lack of SFR-dependence is consistent with the relation 
between metallicity and SFR 
being maintained by episodic metal-poor inflows.
The advent of fresh gas triggers
star formation and drops the metallicity, but it does not
change the pre-existing relative abundance between metals.

\subsection{The stellar mass-metallicity-gas mass  relationship}\label{fmr2}

\citet{2013MNRAS.433.1425B} point out that the FMR
relationship between stellar mass, metallicity, and SFR described in 
Sect.~\ref{fmmr} is just a consequence of a more fundamental relationship
where the SFR is replaced with the HI gas mass.  The authors find the HI-based relationship 
cleaner, and without the saturation at high masses, where irrespectively
of their SFR, all massive galaxies have the same metallicity 
(Fig.~\ref{fmr10}, left panel). A similar anti-correlation between 
HI mass excess and metallicity has been found by 
\citet{2013ApJ...773....4R}. 
\citet{2013A&A...550A.115H} also find the
anti-correlation between metallicity and gas mass at
fixed stellar mass in a  volume-limited sample of 260 nearby 
late-type galaxies. Interestingly,  the relationship is nearly invariant to the 
environment when going from clusters to the field. 
This finding indicates that internal evolutionary processes, rather than 
environmental effects, shape the observed relationships. 

According  to the theoretical models put forward to explain the 
FMR in Sect.~\ref{fmmr}, a relationship between stellar 
mass, metallicity, and gas mass is to be expected. Given the 
FMR, it follows from the KS~relation which states that
the gas mass sets the SFR and vice versa (Eq.~[\ref{kslaw}]). 
Thus the FMR and the relationship 
discussed  in this section are just two renderings of the same 
underlying physical principle, namely, that galaxies are driven out
the equilibrium mass-metallicity relationship  by sudden bursts  
in the inflow rate. These changes increase the HI mass,  
increase the SFR, and decreases the metallicity of the ISM, 
all at once.

\subsection{The mass-metallicity-size relationship}\label{fmr3}

Based on some 44000 SDSS galaxies, \citet{2008ApJ...672L.107E}
found a relation between ${\rm M_\star}$, specific SFR,  and 
galaxy size (as parameterized by the half-light radius). They found 
that at fixed ${\rm M_\star}$, physically smaller galaxies are also 
more metal rich. The metallicity changes by 0.1 dex when the galaxy
size changes by a factor of 2. The authors discard biases 
due to the finite size of the central region used to
estimate metallicities, and to the Hubble type dependence
of the radius. 
A similar relationship between size and metallicity was also found 
at $z\sim 1.4$ by \citet{2012PASJ...64...60Y}.

This non-trivial observational result is, however, a  natural
outcome of the gas accretion driven SF process.
In the stationary state the metallicity is mainly set by the mass 
loading factor ($w$ in Eq.~[\ref{my_metal}]), which changes systematically with 
halo mass, that is to say, with the depth of the gravitational 
potential well from which the baryons have to escape.
The mass-metallicity relationship is interpreted considering that
the halo mass scales with  ${\rm M_\star}$ (Sect.~\ref{fmmr}). 
However, the gravitational
binding energy also depends on the distance of the matter to the
center of the gravitational well. At fixed mass, 
winds escape easier from larger galaxies. Thus they  are expected to have 
larger mass loading factors. If 
the mass loading factor depends  on mass through the escape velocity, 
then it depends as mass/size. A factor 2 increase in size is equivalent 
to a factor 2 decrease in mass, and this scaling is consistent 
with the observed 0.1 dex  metallicity drop when 
either the mass decreases a factor of 2 or the 
effective radius increases by this amount.

\section{Gas accretion and star formation at high redshift}
\label{redshift}

Most of the observational evidence for gas accretion put forward 
so far has been  concentrated on the low redshift universe. 
Theory predicts that  cold gas accretion is particularly important at high redshift, 
when DM haloes  are low mass and the accreted cosmic gas remains cold and 
ready to form stars (Sect.~\ref{theory}).
Thus the high redshift universe corresponds to an epoch of galaxy 
assembly and high SFR.
Here we review observational evidence emphasizing the high redshift aspects 
not addressed elsewhere in the paper.  We include a section on the 
role of mergers, just to complement the statement 
made in Sect.~\ref{introduction} that mergers play only
a subordinate role in galaxy growth outside dense environments.

\subsection{Star formation history of the universe}\label{sfhu}

\begin{figure}
\includegraphics[width=0.8\linewidth]{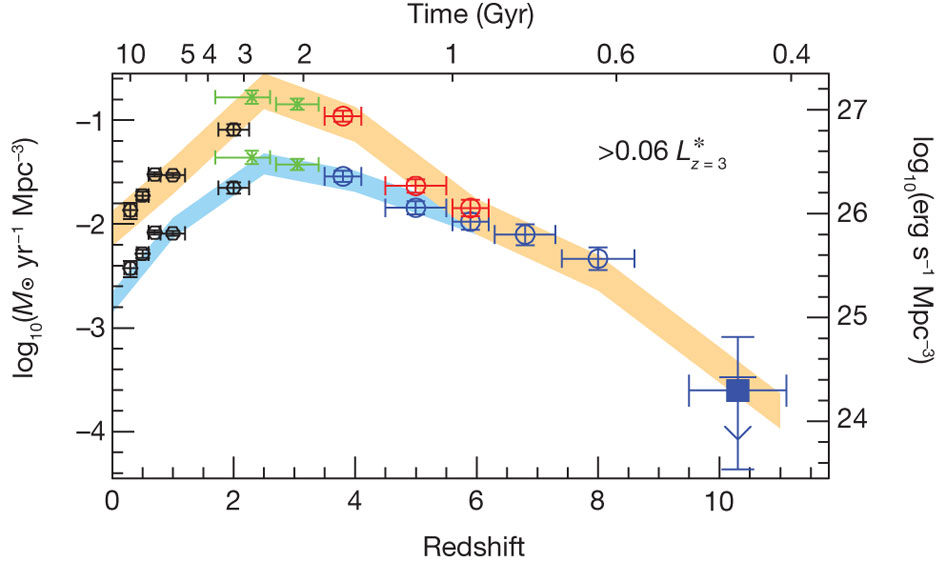}
\caption{
SFR density (dark orange shaded region) and
luminosity density (blue shaded region) in the universe over the last 13.2\,Gyr.
The left  and right axes refer to the SFR and  the luminosity, respectively.
Various symbols correspond to various sources in the literature
as detailed in the paper by \citet{2011Natur.469..504B}  from which
the plot was taken. 
}
\label{bouwens04}
\end{figure}

The star formation rate over the history of the universe has been studied since the mid 90s 
\citep[e.g.,][]{1996ApJ...460L...1L,
1996MNRAS.283.1388M,
1999ApJ...519....1S,
2004ApJ...611..685O,
2014arXiv1403.0007M}.
It requires observing faint objects at high redshift 
and is thus one of the primary  drivers for building larger telescopes 
with adaptive optics and infrared sensitivity. 
Recent observations of the variation of the cosmic SFR density with redshift 
are summarized in Fig.~\ref{bouwens04} \citep[from][]{2011Natur.469..504B}.
This diagram includes a candidate redshift $z\approx 10$ galaxy detected in a dropout 
search in the Hubble Ultra Deep Field. The diagram also includes SFR density 
determinations at  $z\simeq 7$ and $z \simeq 8$ from \citeauthor{2011Natur.469..504B}, 
and values from earlier studies at $z < 4$. Dust corrections are applied,
and are already negligible by $z\simeq 7$ 
\citep{2010ApJ...709L..16O,2010ApJ...709L.133B,2009ApJ...705..936B}. 
Note how the SFR steadily increases with time in the early universe
until reaching a maximum at an age of 2.5 Gyr  ($z\sim$\,3\,-\,2). 

\begin{figure}
\includegraphics[width=0.75\linewidth]{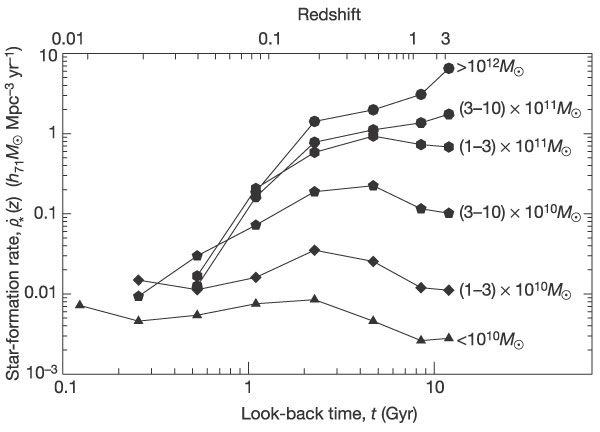}
\caption{
SFR density as a function of 
look-back time (lower x-axis) or redshift (upper x-axis) for galaxies of different
stellar masses.  For clarity, the curves are offset
vertically, successively by 0.5\,dex, except for the most massive galaxies, which are offset by 
an additional 1 dex. Note the clear trend for galaxies with larger present-day stellar mass 
to have formed their stars earlier. The bulk of the SFR at 
$z \simeq 0.5$ comes from galaxies with present-day
stellar masses in the range 3\,-\,$30 \cdot 10^{10}\, {\rm M_\odot}$. 
The figure was taken from \citet{2004Natur.428..625H}. 
}
\label{heavens02}
\end{figure}

In a complementary work, \citet{2004Natur.428..625H} 
analyzed the fossil record of the current stellar populations in nearly $10^5$ nearby galaxies. 
This allowed them to reconstruct the SFR density along the age of the universe 
for galaxies of different masses independently. 
The results are shown in Fig.~\ref{heavens02}.
The inferred star formation history differs according to the (present) galaxy mass so that the 
larger the stellar mass the earlier their stars were formed. High and low mass galaxies have 
very different SF histories. In particular, the global SFR density is driven by the behavior 
of the most massive systems (cf., Figs~\ref{bouwens04} and \ref{heavens02}),
and galaxies with ${\rm M_\star < 3\cdot 10^{10}\,M_\odot}$ today
produced their stars at constant SFR back to at least $z\sim 3$. 

Both the global history (Fig.~\ref{bouwens04}) and the differences among galaxies of different
masses (Fig.~\ref{heavens02}) are well reproduced by  cosmological numerical simulations 
in which most galaxies grow by cold-flow accretion. 
The agreement with the global 
SF history (Fig.~\ref{bouwens04}) can be attributed to  DM halo growth as 
reflected in Eq.~(\ref{eqmdot}):  at very high redshift the haloes grow slowly 
because of their low mass and their shallow gravitational potential. At low redshift 
there is little DM available. The two opposing terms conspire to yield a maximum 
halo growth  at $z\sim$\,2\,-\,3 (Eq.~[\ref{halo}]).
The cosmic star formation history is well reproduced by cosmological numerical simulations
including baryons. For example,
\citet{2010MNRAS.402.1536S} systematically vary details of the modeling to determine the 
dominant physical processes responsible for the SF history.
After a large number of simulations that comprise the  OverWhelmingly Large 
Simulations Project, theoretical diagrams match the observations.
The resulting SFR is limited at high redshift by the build-up of dark matter haloes, 
it reaches a broad maximum at intermediate redshift, and then
 decreases at lower redshift.  The decrease at late times is due to the quenching of SF by lower 
cooling rates in hotter and lower density gas,  in addition to gas exhaustion and self-regulated 
feedback from stars and black holes. 
Similar results are provided by, e.g., 
\citet{1991ApJ...379...52W}, 
\citet{2003MNRAS.341.1253H},
\citet{2008MNRAS.391..481S},  
\citet{2009MNRAS.399.1773C}, and 
\citet{2009MNRAS.393.1595C}.
\citet{2010MNRAS.402.1536S} 
stress that without AGN feedback it is difficult to match the steep decline in the cosmic SFR 
below $z=2$. 
However, it is not yet clear  whether feedback from massive stars can solve the problem. 
The role of stellar feedback through winds and SNe is an active topic of research at this
moment.

Galaxies of different masses follow very different SF histories (Fig.~\ref{heavens02}).
Low mass galaxies have had a more uniform SFR, a fact also well captured
by simulations \citep[e.g.,][]{2009MNRAS.397.1776F,2009MNRAS.393.1127S}.

\subsection{The role of mergers}\label{role_merger}
Gas accretion and feedback, rather than mergers, are the primary processes responsible 
for the evolution of the cosmic SFR density.  This is clear from numerical simulations
which point out that accretion of cold gas is the dominant growth mechanism 
by about an order of magnitude over mergers 
\citep[e.g.,][see also  see Sec.~\ref{introduction}]{2011MNRAS.413.1373W,2013MSAIS..25...45C}.
Observations seem to favor this view too.

Galaxies at $1 < z < 3$ are often clumpy disks 
\citep[e.g.][]{1995MNRAS.275L..19G, 1996MNRAS.279L..47A, 1996ApJS..107....1A, 
1996AJ....112..359V, 1995AJ....110.1576C, 1999ApJ...510...82I, 2004ApJ...603...74E, 
2005ApJ...631...85E, 2007ApJ...658..763E,2009ApJ...701..306E,2009ApJ...706.1364F}
with large SFR in excess of 100\,\smyr\ \citep[e.g.,][]{2011MNRAS.417..289B}.  If mergers were responsible for 
the enhanced star formation, 
observations should show double nuclei and tidal features, 
which are indeed observed in some galaxies but not in these clumpy types \citep{2007ApJ...663..734E}. 
Unfortunately, there is no direct evidence of gas accretion in the clumpy galaxies; in one system at z=1.6, UDF 6462, 
the metallicity map shows an overall value of about 0.5\,$Z_\odot$  with a central concentration and an outward 
decrease \citep{2008A&A...486..741B}. The negative gradient is consistent with 
regular inside-out star 
formation with no excess of metal-poor gas in the inner regions.
Other high-$z$ galaxies show positive gradients, though 
\citep[see Sect.~\ref{hiiregions} and ][]{2012A&A...539A..93Q,2013ApJ...765...48J}.

H$\alpha$ emitters, detected from a flux excess in the Spitzer/IRAC 3.6-$\mu$m band,
comprise more than 70\,\% of the galaxies with redshifts  $3.8 < z < 5$.
\citet{2013ASPC..477..185S} find that the rest-frame UV to optical 
morphologies for half of them do not show any signs of mergers or 
tidal interactions. They have large H$\alpha$ equivalent widths
despite their relatively old stellar population age,  which
suggests that H$\alpha$ emitters are producing stars continuously at a 
constant rate, rather than through multiple bursts invoked by major 
mergers. Their continuous star formation rate, relative 
isolation, and number density are consistent with star formation driven by gas infall. 

\citet{2013MNRAS.430.1051C} find that the observed merger rate alone cannot
explain the mass growth rate of massive galaxies  
$({\rm M_\star > 10^{11}\,M_\odot})$. An additional gas accretion of 
90\,${\rm M_\odot\, yr^{-1}}$ is needed. Such accretion would account for 
36\,\% of the stellar mass growth and  58\,\% of the star formation in the 2\,Gyr 
period from $1.5<z<3$. This holds for the most massive galaxies, but
the need for gas accretion is even more extreme for less massive galaxies. 
Based on HST imaging, \citet{2013MNRAS.429L..40K}   find that only 27\,\%\
of the star formation budget is hosted by major mergers. Discounting the SFR
that even non-interacting galaxies have, the authors
estimate an upper  limit of  15\,\% as the contribution of major mergers  
to star formation in $z\sim2$ massive galaxies 
(${\rm M_\star > 10^{10}\,M_\odot}$). 

MW-like progenitors also grew by gas accretion from $z=2.5$ to the present,
according to \citet{2013ApJ...771L..35V}.  They study the mass assembly
of galaxies expected to have present day ${\rm M_\star \sim 5\cdot 10^{10}\,M_\odot}$.
In marked contrast to the assembly history of massive ellipticals, mass growth 
is not limited to large radii.  The mass growth takes place in a fairly uniform way, with the galaxies 
increasing their mass at  all radii.  After $z = 1$ the growth in the central regions gradually stops and the 
disk continues to be built up. None of these features reflect the dramatic effects expected 
from major mergers. 

Another indication disfavoring mergers comes from the increase of 
tadpole or cometary galaxies in high redshift fields. These galaxies,
with a bright head on a faint tail, exist in the local universe as well as at high 
redshift \citep{1996AJ....112..359V}, but they are much more common
at high redshift where $\sim\,10\,\%$ of the resolved galaxies in the Hubble ultra deep field 
have this  shape \citep{2005ApJ...631...85E}.  Local tadpoles are unusual but, surprisingly, 
they turn out to be very common among extremely low metallicity dwarfs. These and other 
properties indicate that they are disk galaxies in early stages of assembly through gas accretion 
(Sects.~\ref{hiiregions} and \ref{star_accretion}). Low and high redshift tadpoles
 show a  continuum in their properties
\citep[][]{2010ApJ...722.1895E,2012ApJ...750...95E}, therefore, the physics behind them 
is likely to be common. High-$z$ tadpoles may represent disks 
being formed by accretion too. Thus, the large number of tadpoles observed
at high $z$ is consistent with an increase in gas accretion to the detriment 
of mergers. 
Tadpoles are only a particular sub-set within the BCD family. 
The abundance of the BCD high-$z$ counterparts, the clumpy-disk galaxies, also increases   with 
increasing $z$
\citep{2005ApJ...631...85E,2009MNRAS.400.1121S,2011ApJ...733..101G,2012MNRAS.422.3339W}.
Therefore the arguments used for tadpoles supporting gas accretion 
can be extended to BCDs as well.

%
\subsection{The cosmic web in absorption}\label{cwabs}

Star formation in galaxies is self-regulated by the intimate connection 
between gas mass and SFR, 
so that the SFR plus 
outflows driven by feedback tend to balance the cosmic gas accretion rate 
%
%
(Sects.~\ref{stationary_state} and \ref{winds}).
These inflows and outflows controlling galaxy evolution show up as intervening 
absorption systems along the lines-of-sight to background sources, typically QSOs.  
They include DLAs around galaxies (${\rm N_{HI}}>10^{20}$\,cm$^{-2}$)
as well as LLS (${\rm N_{HI}}>10^{17}$\,cm$^{-2}$) and the rest of the Ly$\alpha$ 
forest (${\rm N_{HI}}< 10^{17}$\,cm$^{-2}$)
\citep[e.g.,][]{2005ARA&A..43..861W,2011ApJ...736...48R,2013A&A...552A..77K}. 
Numerical simulations predict  Ly$\alpha$ column densities 
spanning the range from the Ly$\alpha$~forest to
DLAs, inflow velocities $\sim 100$\,\kms , and metallicities 
between $10^{-2.5}$ and $10^{-1}\,Z_\odot$  (Sect.~\ref{prediction_obs}).
All of these features are present in the  observed absorption systems.

DLAs are predominantly neutral gas reservoirs 
with metallicities that decrease with increasing redshift.  
From $0<z<5$, the relationship has a slope of -0.2 dex  per unit $z$,  
a large scatter ($\pm 1$\,dex at a given $z$),  and a lower limit
that extends at least to $Z_\odot/3000$ 
\citep{2012ApJ...755...89R,2014ApJ...782L..29R,2011MNRAS.417.1534C}. 
There are also reports of null detection at high redshift:
$Z < 10^{-4.2}\,Z_\odot$ at $z=3.4$ \citep[][LLS]{2011Sci...334.1245F},
and $Z < 10^{-4}\,Z_\odot$ at $z=7.1$ \citep[][LLS/DLA]{2012Natur.492...79S}.
The $\alpha$/Fe ratios of the most metal-poor DLAs are similar to
those of MW halo stars with metallicities below  $Z_\odot/100$,
suggesting that DLA-like clouds may have been the source of gas
that formed halo stars 
\citep{2011MNRAS.412.1047C,2011MNRAS.417.1534C}.
A similarly weak evolution with redshift of the metal content
of the IGM had been noted by
\citet[][see also \citeauthor{2009MNRAS.395.1476R}~\citeyear{2009MNRAS.395.1476R}]{2003ApJ...594..695P},
who speculated  that the observed metals were created by the sources
responsible for the re-ionization of the universe or,
alternatively, by outflows from star forming galaxies at later times.

\citet{2011ApJ...743..207R} reported  an LLS at $z=0.3$ with very low metallicity gas at
37~kpc from a near solar-metallicity 0.3\,$L_\star$  star-forming galaxy. This was
taken as proof of gas infalling onto the galaxy (see the discussion in
Sect.~\ref{prediction_obs}).  
From this and other evidence from the literature, \citeauthor{2011ApJ...743..207R}
concluded that 50\,\% of the LLSs have such low metallicities (a few percent $Z_\odot$)
and are near  ($< 100$\,kpc) a sub-$L_\star$ galaxy with solar metallicity. 
In a subsequent work by this team, also mentioned in Sect.~\ref{neutral_gas},
\citet{2013ApJ...770..138L} conducted a survey of HI-selected LLSs to provide 
an unbiased metallicity study of the CGM at $z < 1$. 
Their main result is that the metallicity distribution of the LLS is 
bimodal, with one peak at low metallicity ($0.025\,Z_\odot$) and the other
at high metallicity ($0.5\,Z_\odot$).  The metal-poor gas has properties consistent with
cold accretion streams, while the metal-rich gas likely traces winds, 
recycled outflows and tidally stripped gas. The behavior of the bimodal
metallicity is evidence that inflows and outflows do not mix, and
that outflows do not prevent inflows, in agreement with 
numerical simulations (Sect.~\ref{winds}).

Multi-temperature gas in galaxy haloes has been reported too.
\citet{2013ApJ...776L..18C} detected  Ly$\alpha$, deuterium, OI, and other 
metals in the CGM of a $z=2.5$ star-forming galaxy at an impact parameter 
of 58~kpc. Several components are detected, one of which 
has low temperature ($< 2\cdot 10^4$\,K) and metallicity  
($0.01\,Z_\odot$), which is taken as a direct detection of a high 
redshift cold accretion stream. The CGM of this galaxy seems
to be highly inhomogeneous. The majority of the gas is in a cool, metal-poor and 
predominantly neutral phase, but the majority of the metals are in a highly 
ionized phase. 
The multi-phase nature of the haloes is revealed in the study of 
a LLS at redshift $\simeq 0.36$ by  \citet{2011ApJ...733..111T}. 
They detected high and low ionization species.  
The strong detected OVI seems to arise in 
interface material surrounding the photoionized clouds responsible for
the low ionization absorption. 
A cold phase has also been found by \citet{2011ApJ...743...95G}
in an overdensity of galaxies at $z=1.6$, and by \citet{2012MNRAS.427.3029K}
along the projected minor axis of a star-forming spiral galaxy at z = 0.7.
Low temperature is not necessarily equivalent to cold accretion. 
Cold gas inflow has been unambiguously observed towards 6 galaxies 
in the redshift range $0.35<z<1$ by \citet{2012ApJ...747L..26R}. In this
case the metallicity is supposed to be high and so the authors  
infer that the material is not from the cosmic web but from dwarf satellites 
or galactic fountains. 

When several lines-of-sight to QSOs are present near a galaxy, 
the halo gas can begin to be mapped. This is the case  for two intermediate redshift galaxies,
at z=0.48 and 0.78,  which have multiply-lensed QSOs at projected distances  of 50 and 33 kpc, 
respectively \citep{2014MNRAS.438.1435C}. MgII absorption lines are detected along each of the 
4 sightlines in both galaxies. A comparison of the linewidths and velocities with
that of their galaxy disks 
rules out a disk origin or a wind outflow. Instead, the MgII line kinematics are consistent
either with infalling streams from the cosmic web or  from tidally stripped gas.

One of the predictions of cosmic gas accretion refers to very 
massive galaxies and the quenching of star formation. 
If a galaxy becomes too massive, the cosmic web gas is shock-heated 
when entering the halo, and it takes a long time before 
the gas settles down and can be used to form new stars 
(Sect.~\ref{theory} and Sect.~\ref{physical_picture}).
This seems to be happening in an HI absorbing complex at z = 0.7
with an impact parameter of 58 kpc from  
an elliptical galaxy with $10^{13.7}\,{\rm M_\odot}$ virial mass \citep{2012ApJ...760...68C}.
Ionization models suggest a cold gas structure surrounded by a hot cloud. 
The authors interpret the HI complex as a metal-poor filamentary structure
being shock heated as it accretes into the halo of the galaxy. 

A critically important result from cosmic Ly$\alpha$ absorption systems 
is the fact that the average neutral gas density in the universe 
has not changed much during the 
last 10\,Gyr. From ESO UVES spectra of 122 QSOs and data from the 
literature, \citet{2013A&A...556A.141Z} find a constant neutral gas mass 
density in the universe for $0.1<z<5$,  even though star formation varies 
significantly during that period (Fig.~\ref{bouwens04}).  At the peak SFR 
density ($z\sim 2$), the time scale to consume all of the cosmic gas 
in galaxies is only 1\,Gyr, 
therefore, the total neutral gas has to be continuously replenished, partly 
through galaxy outflows but mostly by recombination of ionized gas.
The source of this neutral gas would be the ionized IGM that contains most
of the baryons \citep[e.g.,][]{2004ApJ...616..643F}.

%
\subsection{The cosmic web in emission}\label{cosmic_emission}

The cosmic web gas is also detected in emission as Ly$\alpha$ blobs  
\citep[LABs;][]{
1999AJ....118.2547K,
2000ApJ...532..170S,
2006MNRAS.370.1372F,
2009ApJ...696.1164O}. 
%
Cold-flow streams contain partly ionized  gas undergoing continuous 
recombination and should produce a hydrogen emission line  spectrum
(Sect.~\ref{prediction_obs}). \citet{2010MNRAS.407..613G} determine the 
Ly$\alpha$ flux to be expected from the cold streams that fed galaxies
at high redshift. Model MW haloes at $z=2.5$ have a peak Ly$\alpha$
surface brightness of $2\cdot 10^{-17}$\,erg\,cm$^{-2}$\,s$^{-1}$\,arcsec$^{-2}$,
sizes from 50 to 100\,kpc, and luminosities between $10^{43}$ and 
$10^{44}$\,erg\,s$^{-1}$. These characteristics agree with the LABs,
primarily observed at redshifts $2<z<3$. 
The number of known LABs is still not very large.
For example, a narrow-band imaging survey with the Subaru/Suprime-Cam 
revealed 14 LAB candidates with a mayor axis diameter larger than 100 kpc 
\citep{2011MNRAS.410L..13M}. Their shapes range from circular to filamentary.
The LAB sample shows a possible morphology-density relationship: filamentary
LABs are in average density environment while circular ones reside in both average
and overdense environments. \citeauthor{2011MNRAS.410L..13M} point
out that the more filamentary LABs may relate to cold gas
accretion and the more circular ones may relate to large scale gas outflows.

\begin{figure}
\includegraphics[width=0.9\linewidth]{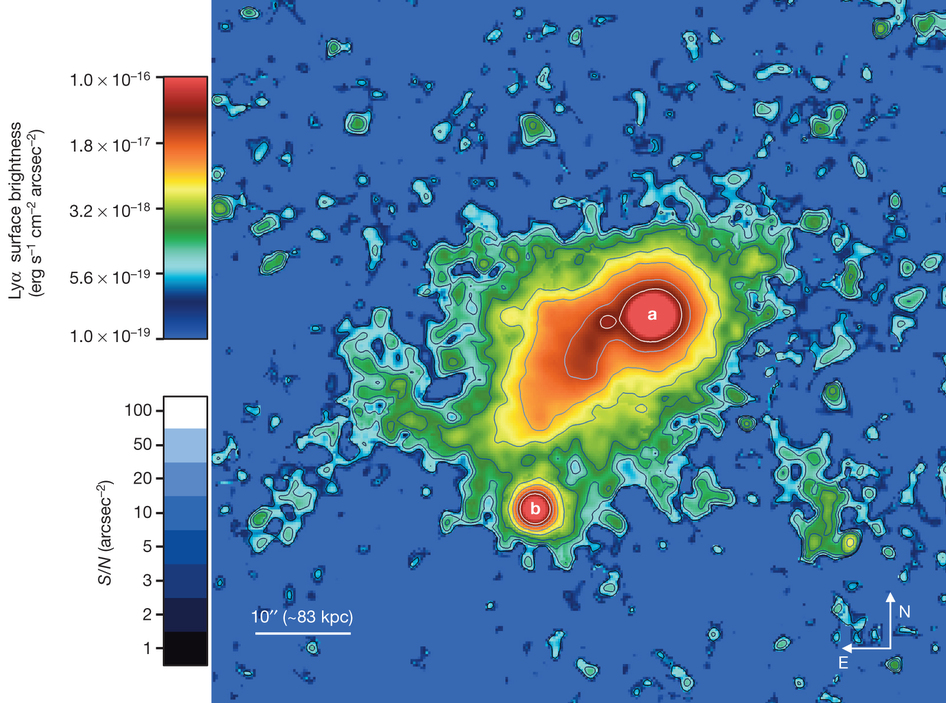}
\caption{
Ly$\alpha$ image of the  UM 287 nebula. The location of the radio quiet QSO UM 287 is labelled with $a$. 
The colour map and the contours indicates, respectively, the Ly$\alpha$ surface brightness (upper colour scale) 
and the signal-to-noise ratio per square arcsec aperture (lower colour scale). The extended emission spans a 
projected angular size of 55\,arcsec or 460\,kpc, which is well beyond the virial radius of any plausible 
associated dark-matter halo and therefore traces intergalactic gas.
The object marked with $b$ is an optically faint QSO at the same redshift $z\sim 2.3$. 
The large distance between the two QSOs and the very broad morphology of the nebula 
argue against the possibility that it originates from an interaction between the two QSO 
host galaxies. The work and the figure are from \citet{2014Natur.506...63C}.
}
\label{nature_cantalupo}
\end{figure}

The cosmic web gas also emits through fluorescence of Ly$\alpha$ 
photons originally produced by a relatively nearby source such as a powerful 
starburst galaxy  or a luminous QSO \citep{2012MNRAS.425.1992C}.
Figure~\ref{nature_cantalupo} shows  the Ly$\alpha$ image 
of the UM287 nebula at $z=2.3$ taken from \citet{2014Natur.506...63C}.
In the figure the nebula appears broadly filamentary and asymmetric, extending 
mostly on the eastern side of QSO $a$ up to a projected distance 
of about 55 arcsec or 460  kpc, which is well beyond 
the virial radius of any plausibly associated dark-matter halo.
Therefore the nebula traces intergalactic gas. 
Fluorescent Ly$\alpha$ photons have also been observed 
by \citet{2011MNRAS.418.1115R} as an elongated structure at $z=3.3$ with a peak 
surface brightness around $10^{-17}$\,erg\,cm$^{-2}$\,s$^{-1}$\,arcsec$^{-2}$.
Several of these Ly$\alpha$ emitting structures without continuum counterparts 
have been found by  \citet{2012MNRAS.425.1992C} near a 
hyperluminous QSO at $z=2.4$. 

\section{Fraction of the star formation sustained
by cosmic gas accretion}\label{myfraction}

So far we have been citing evidence for gas accretion 
driving SF. The question arises as to what is the actual 
contribution of this cosmic web gas to the observed SF; that is,
what is the fraction of SFR in gas that has not previously been inside other stars 
compared  to SFR in gas that has been reprocessed in 
the galaxy? 
From a theoretical  point of  view, the fraction of recently accreted
gas involved in star formation must be  very high. 
Equation~(\ref{cosmosfr}) shows that in a quasi-stationary state more than 
half of the gas mass participating in any typical star formation event 
comes directly from cosmological infall;
it is not gas ejected from previous star-formation episodes in the
galaxy.
Theory also tells us that the fraction depends very much on the galaxy 
tendency to lose mass through winds, so that the larger the losses the 
higher  the fraction of cosmological gas needed to maintain the SF. 
The fraction reaches 90\,\%  for a  mass loading factor $w=5$  typical of  a
galaxy with ${\rm M_\star\simeq 10^9\,M_\odot}$ (see Sect.~\ref{winds}). 
This section collects a few estimates of this fraction, i.e., of 
the percentage of star formation sustained by gas accretion.
The list is not exhaustive but the number is limited mainly 
because there are only a few estimates of this parameter in the 
literature. In general, observations support the theoretical expectations, 
but the issue is far from settled.

\begin{figure}
\includegraphics[width=0.6\linewidth]{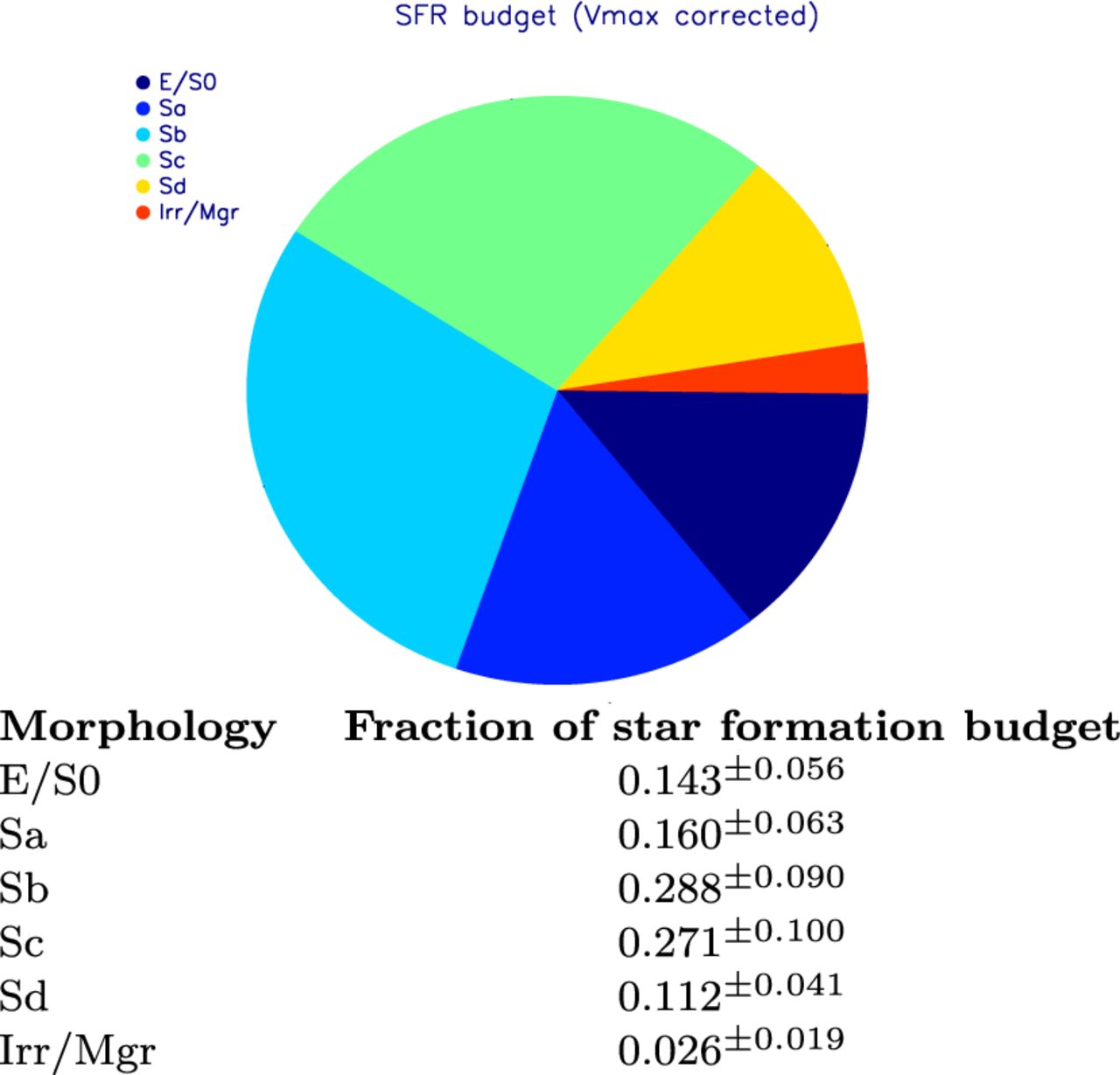}
\caption{Cosmic star formation budged at $z < 0.07$ split by 
morphological type. It is corrected for Malmquist bias, therefore,
they represent the SF in a fixed volume. 
Early-type galaxies (E/S0) account for 14\,\% of the SFR although they 
contain 50\,\% of the stellar mass.  
Taken from \citet{2014MNRAS.437L..41K}.
\label{kaviraj}
}
\end{figure}

\citet{2014MNRAS.437L..41K} estimate the contribution of 
minor mergers to the cosmic SF budget but, in our context,  minor
mergers are equivalent to  gas accretion events.
Neither observations nor numerical simulations 
can easily distinguish a gas rich merger from a starless gas 
clump  (Sect.~\ref{theory}). \citeauthor{2014MNRAS.437L..41K}
computes the fraction of the SFR associated with each Hubble type using
a sample of 6500 SDSS-Stripe~82 galaxies. He assumes that 
all SF in early types is due to an external gas supply, 
because most of their stars were formed so long ago than the
debris from those stars render very little gas at this moment. 
Early types provide only 14\,\%\ of the current SFR despite the fact that
they account for 50\,\%\ of the stellar mass (Fig.~\ref{kaviraj}).
If two galaxies have the same mass and the same environment, then
they should have the same cosmic gas accretion rate, independently 
of their Hubble type. Under this premise,  
he finds  
that  at least 21\,\%  of the SF activity in late types 
is due to accretion. Adding up the contributions from early and
late types, he concludes that at least  35\,\% of the star formation 
is due to external accretion. This figure represents  a lower limit 
since it assumes that  early and late types have available similar 
gas supplies, which underestimates late type gas resources.  

%
\citet{2013MNRAS.430.1051C} study the mass growth of massive
galaxies ${\rm M_\star > 10^{11} M_\odot}$ at  $1.5 < z < 3$.
They estimate  the evolution of stellar mass in these systems 
from the observed SFRs and the amount of stellar and gas mass 
added due to observed major and minor mergers.
The measured gas mass is insufficient to maintain the inferred star 
formation history, and the needed additional gas mass cannot be accounted 
for by gas delivered through minor and major mergers
or by recycling of stellar ejecta. The estimates show that during this epoch, and for 
these massive galaxies, 50\,\%  of the baryonic mass assembly  results from gas 
accretion and  unresolved mergers, 25\,\%  is put into place through existing stars 
from mergers, with the remainder of the mass in the form of
gas brought in with these mergers. 
As for SF during this epoch, 66\,\%  is the result of gas accretion. 

%
\citet{2012ApJ...750..142B} model the FMR  (Sect.~\ref{fmmr})
assuming that  external gas infall produces  the observed star formation. 
All galaxies with  ${\rm M_\star < 2\cdot 10^{10}\,M_\odot}$ and many with 
masses above this threshold follow the model. 
They conclude that most of the star-forming galaxies  with 
stellar masses  ${\rm M_\star  < 2 \cdot  10^{10}\,M_\odot}$, 
and many with ${\rm M_\star  > 2 \cdot  10^{10}\,M_\odot}$ 
appear to be fed by low-metallicity gas infall.


\section{What we understand, and what we do not, 
and pathways to follow up}\label{conclusions}

Numerical simulations predict that the accretion of metal-poor gas from 
the cosmic web drives the growth of disk galaxies. They describe
galaxies as open systems where gas inflows and outflows determine the 
gas available to form stars and therefore the global properties
of the stellar populations. Observational evidence for 
gas accretion in galaxies is also numerous although indirect.
The gas that falls in or goes out is tenuous, patchy, partly  
ionized, multi-temperature, and large scale;  therefore, it is
hard to show in a single observation. A multi-wavelength 
multi-technique approach is often needed to formulate 
a compelling case for gas flows associated with SF.
Among the many examples given in this paper, we have selected 
three cases for the sake of illustration:
(1) the short gas consumption time-scale compared to the 
long SF record in most galaxies (Sect.~\ref{neutral_gas}),
(2) the metallicity-SFR relationship, so that
at a given stellar mass, the larger the current SFR 
the lower the metallicity of the gas producing stars (Sect.~\ref{fmmr}),
and (3) the large amount of gas 
hundreds of kpc away from galaxies but with properties related to
the closest neighboring galaxy  (Sects.~\ref{neutral_gas} and \ref{cwabs}).

The global picture of the interplay between cosmic gas accretion 
and star formation seems to be understood at a basic level (Sect.~\ref{theory}).   
In the long term, the gas accretion rate has to balance the
SFR since the gas available to form stars is consumed  quickly.
This balance is self-adjusted by the galaxy 
modifying the  mass of gas available to form stars which,
through the KS relation, sets the SFR. The details of the process 
remain to be worked out, though. They  depend on the coupling between 
physical processes that involve completely different physical scales: 
from cosmic web structures to molecular clouds and AGN engines.

Many ingredients needed to complete the observational
picture remain to be gathered and understood. For example, 
the large-scale  of the gas in the cosmic web should be imaged.
Emission has been detected with 10-m class telescopes 
in Ly$\alpha$ at high redshift, but it  may also be observable 
in H$\alpha$ in the local universe.
Using SFR as a proxy for gas accretion rate, star-forming galaxies 
can be used to map the gas accretion rate in the universe, near and far. 
The FMR provides a diagnostic potential that has not been 
fully exploited yet. Assuming that it is set by cosmic gas infall 
triggering SF  (Sect.~\ref{fmmr}), it allows us to measure various key 
parameters in galaxy formation including mass loading factors and 
metallicity of the accreted gas. This is a venue to be explored very
much in the vein of the work by  \citet{2013arXiv1311.1509F}. 
The neutral gas inflow rate directly observed as HVCs in the MW
is one order of magnitude smaller than that required to maintain the 
SFR (Sect.~\ref{neutral_gas}).  This poses a  problem. Several possibilities have been explored 
(e.g., \citeauthor{2008A&ARv..15..189S}~\citeyear{2008A&ARv..15..189S};
\citeauthor{2012ApJ...759..137J}~\citeyear{2012ApJ...759..137J})
but the issue  has not been settled yet.
Details of the interaction between a cold metal-poor gas
stream and a galaxy disk remain to be modeled. 
Predictions on how this interplay excites starbursts and
their metallicities are needed  to secure our
understand of the FMR (Sect.~\ref{fmmr}).  
Outflows from dwarfs seem to be  the main source of metals in the IGM. 
However the mixing processes of these outflows with the IGM are poorly 
captured by numerical simulations.
The cooling of halo gas onto supernova debris in galactic fountains 
has also been suggested as a source of the metal-poor gas
driving SF in the MW and other galaxies  (Sect.~\ref{winds}).
Can this possibility be proved or disproved? 
Cosmic web gas might also be detected through the Sunyaev-Zeldovich effect
imprinted on the Cosmic Microwave Background; this  is a promising technique 
under development now \citep[e.g.,][and references therein]{2013MNRAS.432.2480G}.

These and other future research programs will contribute to our understanding 
of cosmic accretion.  They will eventually complete the picture that has been  emerging 
over the last few years, in which cosmic gas accretion plays a fundamental role 
in sustaining star formation in disk galaxies. 

\begin{table}
\caption{List of main acronyms and symbols defined and used along the text}
\label{acronyms}      
\begin{tabular}{ll}
\hline\noalign{\smallskip}
Acronym & Expansion \\
\noalign{\smallskip}\hline\noalign{\smallskip}
AGN&Active galactic nucleus\\
BCD &Blue compact dwarf\\
BH &Black hole\\
CGM &Circum galactic medium\\
DM &Dark matter\\
DLA&Damped Lyman-$\alpha$ absorbers\\
FMR&Fundamental metallicity relationship\\
HVC&High velocity clouds\\
IGM&Intergalactic medium\\
ISM&Interstellar medium\\
IMF&Initial mass function\\
KS&Kennicutt-Schmidt\\
${\rm \Lambda CDM}$& ${\rm \Lambda}$ cold dark matter\\
LAB&Lyman $\alpha$ blobs\\
LLS&Lyman limit systems\\
${\rm M}_\star$& Stellar mass\\
${\rm M_g}$& Gas mass\\
${\rm M_{halo}}$&Dark matter halo mass\\
MW&Milky Way\\
QBCD &Quiescent blue compact dwarf\\
QSO&Quasar\\
SF &Star formation\\
SFR &Star formation rate\\
SN, SNe& Supernova, Supernovae\\
SPH&Smoothed particle hydrodynamics\\
UV&Ultraviolet\\
$w$&Mass loading factor\\
XMP& Extremely metal poor\\
$z$&Redshift\\
$Z$ &Metallicity\\
\noalign{\smallskip}\hline
\end{tabular}
\end{table}

\begin{acknowledgements}
Daniel Ceverino helped us in dissecting the predictions
of the cosmological numerical simulations, 
and Sects.~\ref{physical_picture} and \ref{prediction_obs} 
owe much to his patience.  
Discussions with  Claudio Dalla Vecchia 
were also fundamental for the development of many aspects 
of the work. 
Thanks are due to Mercedes Filho for discussions and clarifications
on the HI content of galaxies, and 
%
%
to 
Practika Dayal, Martin Harwit, and Simon Lilly 
for discussions on modeling the FMR.
Fernando Atrio pointed out the work to
image the cosmic web through the 
Sunyaev-Zeldovich effect.
We are grateful to Joss Bland-Hawthorn, 
Filippo Fraternali, Jairo M\'endez-Abreu, and Max Pettini 
for comments on the manuscript.
We acknowledge the intensive use of the 
NASA Astrophysics Data System (ADS), including 
{\em Private Libraries}.
The work has been partly funded by the 
Spanish MINECO, project AYA~2010-21887-C04-04 and
Severo Ochoa program SEV-2011-0187.
Figures~\ref{tadpole} and \ref{lopsideness} have been reproduced with permission 
of the AAS, and Figs. \ref{izw18} and \ref{polar_ring} with permission of ESO.  
\end{acknowledgements}

%
\newcommand\aj{AJ}
\newcommand\apj{ApJ}
\newcommand\apjl{ApJ}
\newcommand\apjs{ApJS}
\newcommand\mnras{MNRAS}
\newcommand\aapr{A\&ARev}
\newcommand\araa{ARA\&A}
\newcommand\aap{A\&A}
\newcommand\nat{Nature}
\newcommand\pasp{PASP}
\newcommand\pasj{PASJ}
\newcommand\nar{NewAR}
\newcommand\na{NewA}
\newcommand\aaps{A\&AS}
\newcommand\apss{Ap\&SS}
\newcommand\physrep{Physics Reports}
\newcommand\bain{BAIN}
\bibliographystyle{aa}

%

%

\end{document}